\documentclass{aa}  % REMEMBER TO CHANGE FIGURE SIZES!

\usepackage{graphicx}
\usepackage{txfonts}
\begin{document}

   \title{Energetic proton back-precipitation onto the solar atmosphere in relation to long-duration gamma-ray flares}

%   \subtitle{Back-Precipitation onto the Solar Atmosphere}
\titlerunning{Back-precipitation onto the solar atmosphere}
\authorrunning{Hutchinson et. al.}

   \author{A. Hutchinson
          \inst{1}
          \and
          S. Dalla\inst{1}
          \and
          T. Laitinen\inst{1}
          \and
          G. A. de Nolfo\inst{2}
          \and
          A. Bruno\inst{2,3}
          \and
          J. M. Ryan\inst{4}
          \and
          C. O. G. Waterfall\inst{1}
          }

   \institute{Jeremiah Horrocks Institute, University of Central Lancashire, Preston, PR1 2HE, UK\\
                \email{AHutchinson3@uclan.ac.uk}
        \and
             Heliophysics Division, NASA Goddard Space Flight Center, Greenbelt, MD, USA\\
         \and
            Department of Physics, Catholic University of America, Washington DC, USA\\
        \and
            Department of Physics and Astronomy, University of New Hampshire, Durham, NH, USA\\
              }

   \date{}

% \abstract{}{}{}{}{} 
% 5 {} token are mandatory
 
  \abstract
   {Gamma-ray emission during long-duration gamma-ray flare (LDGRF) events is thought to be caused mainly by $>$300 MeV protons interacting with the ambient plasma at or near the photosphere. Prolonged periods of the gamma-ray emission have prompted the suggestion that the source of the energetic protons is acceleration at a coronal mass ejection (CME)-driven shock, followed by particle back-precipitation onto the solar atmosphere over extended times.}
   {We study the latter hypothesis using test particle simulations, which allow us to investigate  whether scattering associated with turbulence aids particles in overcoming the effect of magnetic mirroring, which impedes back-precipitation by reflecting particles as they travel sunwards.}
   {The instantaneous precipitation fraction, $P$, the proportion of protons that successfully precipitate for injection at a fixed height, $r_i$, is studied as a function of scattering mean free path, $\lambda$ and $r_i$. Upper limits to the total precipitation fraction, $\overline{P}$, were calculated for eight LDGRF events for moderate scattering conditions ($\lambda$=0.1 au).}
   { We find that the presence of scattering helps back-precipitation compared to the scatter-free case, although at very low $\lambda$ values outward convection with the solar wind ultimately dominates. For eight LDGRF events, due to strong mirroring, $\overline{P}$ is very small, between 0.56 and 0.93\% even in the presence of scattering.}
   {Time-extended acceleration and large total precipitation fractions, as seen in the observations, cannot be reconciled for a moving shock source according to our simulations. Therefore, it is not possible to obtain both long duration $\gamma$ ray emission and efficient precipitation within this scenario. These results challenge the CME shock source scenario as the main mechanism for $\gamma$ ray production in LDGRFs.}
%31(1-(1-(r_i^-1.495/rp^-2))0.5)
%1.450-0.058\% 5-70 solar radii lambda - 0.1 AU

   \maketitle
%
%-------------------------------------------------------------------

\section{Introduction} \label{sec:intro}

The production of solar $\gamma$ rays over extended durations, including times when flare emission in other wavelengths is no longer present, has been observed for decades in association with large energy release events at the Sun (\cite{Rya2000} and references therein). In recent years, however, new data at photon energies $>$100 MeV from the \textit{Fermi} Large Area Telescope (LAT) have shown that these long-duration gamma-ray flares (LDGRFs) are not as rare as previously thought, reigniting debate over their origin (\cite{Ajello_2014,Pes2015,Ack2017,Kah2018,Kle2018,Omo2018,Sha2018,Den2019} and references therein). In these events,  $\gamma$ rays are thought to be generated when $>$300 MeV protons and $>$200 MeV/nuc $\alpha$ particles collide with plasma near the solar surface (1 R$_\odot$) to produce pions that subsequently decay  (e.g. \cite{Sha2018} and references therein). 

%The prolonged $\gamma$ ray emission during LDGRF events is also referred to as Late-Phase Gamma-Ray Emission (LPGRE) \citep{Sha2018} and Sustained Gamma-Ray Emission (SGRE) \citep{Kah2018}.

The presence of LDGRFs imply that highly energetic protons and $\alpha$ particles strike the photosphere over extended time periods, of the order of hours and up to about 20 hours. A number of possible theories for the phenomenon have been put forward. The main two are: a) trapping of flare-accelerated ions within large coronal loops, with the possibility of time-extended acceleration within them \citep{Man1992, Ryan_and_Lee_1991}, and b) time-extended acceleration at a propagating coronal mass ejection (CME) driven shock followed by back-precipitation onto the solar atmosphere \citep{Cli1993}. The $\gamma$ ray emission in these scenarios has been referred to as late-phase gamma-ray emission \citep{Sha2018} and sustained gamma-ray emission \citep{Kah2018}.

\cite{Sha2018} and \cite{Win2018} analysed the association of LDGRFs with soft-X-ray flares, CMEs and near-Earth solar energetic particle (SEP) events and concluded that their most likely origin is back-precipitation after acceleration at a CME-driven shock. %This scenario is also supported by the analysis of the energy released from magnetic reconnection in flaring active regions \citep{Kah2018}. %by correlation studies with type II radio burst data \citep{Gop2018} and
{We refer to this in the following as the CME shock scenario.}
\cite{Jin_2018} studied the 2014-Sep-01 behind-the-limb LDGRF event, performing a detailed analysis of the CME-driven shock up to the time when it reached $\sim 10$ R$_\odot$, focussing on the evolution of its parameters and magnetic connectivity. They find that the compression ratio of their simulated shock displays a similar evolution to the observed $\gamma$ ray profile for the first $\sim 20$ minutes of the event.
On the other hand, a study of a variety of electromagnetic emissions for the same event led \cite{Gre2018} to favour gamma ray production via flare-accelerated protons that remain trapped in large flare loops, explaining the first $\sim 20$ minutes of the Fermi and hard X-ray observations.%On the other hand, a study of a variety of electromagnetic emissions for \textbf{the same} event has concluded that reacceleration of particles trapped on closed coronal loops is the most likely explanation for the observed $\gamma$ ray signatures \citep{Gre2018}.

It has been established that flare acceleration over long timescales is not the source mechanism of LDGRFs. For instance, \cite{Kah2018} used observations of the reconnection rates of flare ribbons and determined that the reconnection episodes do not take place long enough to explain the time-extended $\gamma$ ray emission. \cite{Kle2018} have shown that typical signatures of flare acceleration are not present over long durations in these events.

The back-precipitation scenario has been studied by modelling both the shock acceleration and propagation to the solar surface of the energetic protons. \cite{Koc2015}  concluded that the mechanism is a viable explanation for LDGRFs but pointed out that only about 1\% of the particles accelerated at the shock back-precipitate to the required height. \cite{Afa2018} used a model with strong scattering in the region behind the shock to show that a number of protons sufficient to produce the observed gamma-ray emission (or in some cases considerably more) propagate to the solar surface. \cite{Jin_2018} suggested that scattering associated with turbulence would facilitate particle back-precipitation.
\cite{Kou2020} recently presented a study of the 2017 September 10 event in which they modelled the parameters of the associated CME-driven shock. They conclude that the evolution of the shock and its orientation can explain the time history of the $\gamma$ ray emission observed by \textit{Fermi} LAT, providing further evidence for the CME-driven shock scenario. 

%They concluded that they can explain the $\gamma$ ray emission observed by \textit{Fermi} LAT for the event by considering the evolution of the shock parameters and geometry, indicating the plausibility of the CME-driven shock origin of the $> 300$ MeV protons.

\cite{Hud2018} and \cite{Kle2018} pointed out that magnetic mirroring is a major obstacle for proton back-propagation from a CME-driven shock to heliocentric distances $r \sim 1 \: \textrm{R}_\odot$, where $\gamma$ ray emission takes place. Due to the solar wind expansion and associated $1/r^2$ dependence of the magnetic field magnitude, in the absence of scattering only particles in a very narrow range of pitch angles (the so-called loss cone) are able to avoid reflection as they move towards the Sun. An open question is how this picture is modified by scattering associated with magnetic field turbulence.

\cite{Den2019} analysed 1 AU SEP data from the Payload for Antimatter Matter Exploration and Light-nuclei Astrophysics (\textit{PAMELA}) space experiment, the Geostationary Operational Environmental Satellites (\textit{GOES}) and the twin Solar TErrestrial RElations Observatory (\textit{STEREO}) spacecraft to reconstruct the SEP spatial distribution, accounting for both longitudinal and latitudinal magnetic connectivity, to derive the overall number of protons at 1 AU, $N_\mathit{SEP}$, for 14 events associated with LDGRFs. They compared $N_\mathit{SEP}$ with the number of interacting $\gamma$ ray-producing protons at the Sun, $N_\mathit{LDGRF}$, as inferred from \textit{Fermi}/LAT data by \cite{Sha2018}. They found no correlation between the two populations and showed that in several events $N_{LDGRF} \gtrsim N_\mathit{SEP}$, implying that back-precipitation of a very large fraction of the energetic particles would be required to explain the events within the CME shock acceleration scenario.

Particles accelerated at a CME-driven shock would need to traverse three distinct magnetic field regions in order to back-precipitate: interplanetary space characterised by a magnetic field that can be approximated as a Parker spiral; the corona, with a complex magnetic field configuration consisting of open and closed magnetic field, and the near photosphere region, typically described as a `canopy' \citep{Seck_1991}.

The highest-energy particles are thought to be accelerated around the shock nose, where the shock is strongest and fastest, while its compression ratio and speed decrease quickly at the flanks resulting in a reduced acceleration efficiency (e.g. \cite{Cane_1988,Reames_2009,Hu_2017}). However, other studies have suggested that the shock flanks may contribute (e.g. \cite{Kah_2016}). In general, the position on the shock where the highest-energy particles are accelerated may vary with time and may be strongly influenced by local conditions including the magnetic-field configuration (e.g. \cite{Afa2018,Kong_2019}).

 In this paper we focussed on the CME shock scenario for the origin of LDGRFs and study the back-precipitation of energetic protons towards the solar surface from CME shock heights.
We simulated proton transport by means of a full-orbit test particle code and  address the question of whether particle scattering associated with turbulence may help back-propagation significantly compared to a scatter-free case, analysing the dependence of precipitation fractions on scattering mean free path and injection height.
In this initial study, we used a magnetic field model including only the interplanetary magnetic field (IMF), which as a first approximation is given by a Parker spiral expression.
This is reasonable as a first approximation since in many LDGRF events the CME shock is located beyond source surface heights for most of the duration of the $\gamma$ ray emission. 
Since the magnetic field undergoes an expansion both near the photosphere and within the corona, precipitation fractions from a model including just the IMF will provide upper limits to actual values.

In addition to investigating the general back-precipitation process in an idealised situation, we considered eight specific LDGRF events.  We used information about the CMEs associated with the events and the results of our simulations to obtain upper limits to the total precipitation fraction for the events within the CME shock scenario, in the presence of moderate scattering.

In Section \ref{sec:analytical_expression} a description of our model and its assumptions are given and mirror point radii in the presence of scattering are discussed. The effects of scattering on proton precipitation for instantaneous injections are discussed in Section \ref{sec:Results} and the spatial patterns of precipitation onto the solar surface in Section \ref{subsec:long_lat_pos}. In Section \ref{sec:shock_heights} we analyse shock heights as a function of time for the CMEs observed during eight LDGRF events and in Section \ref{subsec:indiv_event} we then estimate upper limits to their total precipitation fractions. Time profiles of precipitation are discussed in Section  \ref{sec:Timeprofiles} and Section \ref{sec:Disc&conc} presents discussion and conclusions.

\section{Modelling particle back-precipitation}\label{sec:analytical_expression}

 A full test particle model of back-propagation from CME shock heights to the solar surface would require a model of the IMF,  the coronal field (via a potential field source surface or MHD model) and the magnetic field close to the photosphere.
  Analysis of shock heights at times of $\gamma$ ray emission for LDGRF events  (Section \ref{sec:shock_heights}) reveals that, within the CME shock scenario, a very large part of the back-precipitation of energetic particles takes place when the source is in interplanetary space.
 For this reason, in this initial investigation we focussed on the role of the IMF and we considered the Parker spiral as a first approximation. We note that the actual IMF may differ from the nominal Parker spiral and it is known that in some relativistic solar particle events earlier CMEs altered the spiral structure (e.g. \cite{Masson_2012}).

We modelled the propagation of energetic protons towards the solar surface using a full-orbit test particle code, specifically an adapted version of the code used by \cite{Marsh_2013}, with a Parker spiral field given by 
\begin{equation}
    \mathbf{B} = \frac{B_0 r_0^2}{r^2} \mathbf{\hat{e}_r} - \frac{B_0 r_0^2 \Omega \sin{\theta}}{v_{sw} r} \mathbf{\hat{e}_\phi},
    \label{parker_field}
\end{equation}
where $B_0$ is the magnetic field strength at a fixed reference radial distance $r_0$ (we choose $r_0 \: = \: 1$ R$_\odot$), $r$ is the radial distance from the centre of the Sun, $\theta$ is the colatitude, $\Omega$ is the sidereal solar rotation rate, and $v_{sw}$ is the solar wind speed.
Equation \ref{parker_field} is known to be a good approximation down to $r \sim 2.5$ R$_\odot$ (the nominal source surface), below which more complex coronal and photospheric magnetic field are present \citep{Owe_2013}. 
In the simulations we considered a unipolar field with positive magnetic polarity (i.e. outward pointing) using the same parameters as in \cite{Marsh_2013}, which assumes a constant solar wind speed of $v_{sw} = 500 \: \textrm{km s}^{-1}$. Hence pitch angles in the range of $90^\circ  <   \alpha  \leq  180^\circ$ correspond to sunwards propagating particles.

We simulated a 300 MeV mono-energetic proton population, instantaneously injected into a $8^\circ \times 8^\circ$ region, in longitude and latitude, centred on $0^\circ \times 0^\circ$ at a user specified radial height (the injection radius, $r_i$) from the centre of the Sun. We note that acceleration at the shock was not modelled and the shock was transparent to propagating particles, such that propagating particles' trajectories are not affected if they return to the position of the shock. The protons' positions within the injection region are random and the population was isotropic in velocity space at the initial time. We used a small injection region that models only a small portion of the shock. However, this small region placed at different heights can model different parts of the shock thereby allowing descriptions of different possible acceleration sites to be considered (i.e. acceleration at the flanks of the shock or nearer the nose depending on the radial height of the injection region). The equation of motion (see \cite{Marsh_2013} equation 3) for each proton was integrated to determine its trajectory. %Protons are propagated down to 1 R$_\odot$.%or 2.5 

%If a negative polarity was used the protons would gyrate in the opposite direction and drifts would be in the opposite direction.

The effect of magnetic turbulence was described as pitch angle scattering. The intensity of the scattering was determined by the mean free path, $\lambda$. In our simulations scattering events were Poisson distributed temporally with the average scattering time defined by $t_{scat} = \lambda / v$, where $v$ is the particle's initial velocity \citep{Marsh_2013}. During a scattering event the particle's velocity vector is randomly reassigned to another point on a sphere in velocity space (a full discussion of the scattering model can be found in \cite{Dalla_2020}), thereby changing its direction of motion. Each simulation propagated particles through the magnetic field considering a constant $\lambda$.
%$\lambda$ is the parallel scattering mean free path and
To investigate the possible shock source scenario, we ran a number of simulations that model the propagation of protons over 24 hours and involved injection regions located at radial positions in the range $r_i = 5$ R$_\odot$ to $r_i = 70 $ R$_\odot$ and scattering conditions in the range $\lambda$ = 0.0025 AU to $\lambda$ = 1.0 AU. %We follow particles for $t = 24$ hours.

Where protons interact on the Sun to produce $\gamma$ rays is dependent on the local plasma density. It is generally considered that the protons would interact in the lower chromosphere or upper photosphere \citep{Sha2018,Win2018}. However, there have been studies that assumed the protons would interact at greater depths, such as the \cite{Seck_1991} study which assumed they would interact at a depth of 500 km below photosphere. In this work we have neglected this range of interaction heights as they were negligible when compared with the vast interplanetary distances that the protons must propagate through and we assume that pion production takes place at $\sim 1 \: \textrm{R}_\odot$. The test particle code records the time and the associated particle parameters when a particle crosses the 1 R$_\odot$ boundary from larger distances. Particles that cross this boundary were regarded as absorbed and were no longer propagated. We also assumed that all particles that reach $1 \: \textrm{R}_\odot$ go on to produce $\gamma$ rays. %middle right on page 656, they consider column density and inelastic proton-proton cross section in seckel et al paper 

\subsection{Scatter-free mirror point radius}
As charged particles propagate towards the Sun into a region of greater magnetic field strength, their pitch angle is shifted towards $90^\circ$ due to the magnetic mirror effect. The particle's motion is slowed in the direction parallel to the magnetic field line and the component of velocity perpendicular to the field, $v_\perp$, increases. At the mirror point the  component of velocity parallel to the field, $v_\parallel$, goes to zero and the direction of motion reverses.

For a Parker spiral magnetic field, (see Equation \ref{parker_field}) and scatter-free particle propagation, it is possible to derive an expression for the radial position of the mirror point, $r_\mathit{mp}$, analytically given by
\begin{equation}
  r_\mathit{mp} = \frac{B_0}{B_i} \, r_0  \sin{\alpha_i} \left [ \frac{r_0^2  \sin^2{\alpha_i}}{2 \,a^2} + \sqrt{ \frac{r_0^4  \sin^4{\alpha_i}}{4 \, a^4} + \frac{B_i^2}{B_0^2} } \:\right ]^{1/2}_,
  \label{r_mp-B-ratio-in-text}
\end{equation}
where $B_i$ is the magnetic field strength at the starting location of the particle, at radial distance $r_i$, $\alpha_i$ is the initial pitch angle, and $a$ is a function of initial colatitude $\theta$ given by $a = v_{sw}/(\Omega \, \sin{\theta})$.

We note that all terms in Equation \ref{r_mp-B-ratio-in-text} involving the strength of the magnetic field are ratios at two different radial positions. Hence the mirror point radius under scatter-free conditions is independent of the magnetic field strength and depends only on the $B_i / B_0$ ratio.

\subsection{Mirror point radius in the presence of scattering}\label{subsect:mirror_protons}
\begin{figure}[h] %width = 1.0\linewidth, height = 7cm %scale = 0.3
    \centering
    \includegraphics[keepaspectratio = true,width = 1.0\linewidth ]{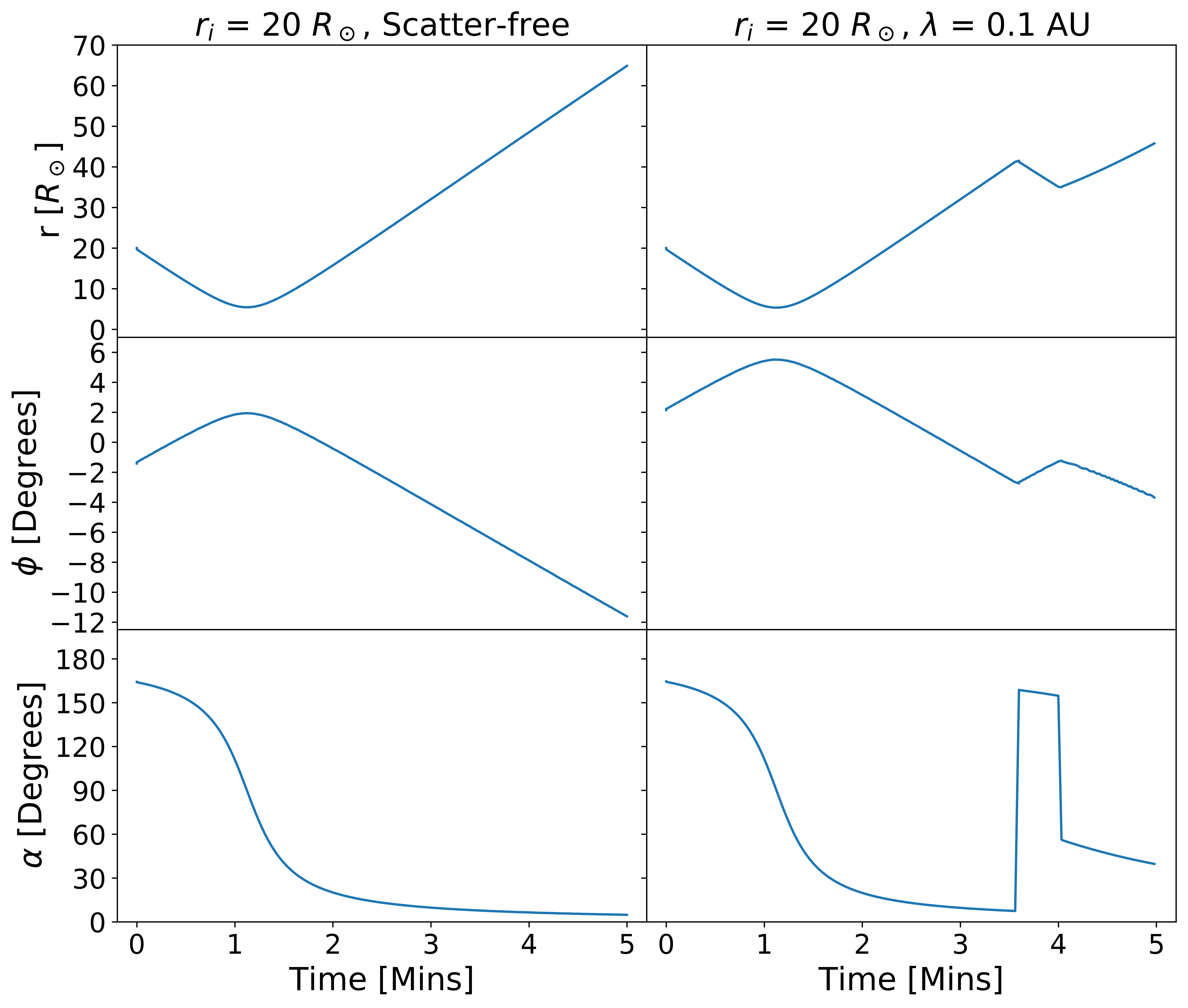}%[keepaspectratio = true,width = 1.0\linewidth ] [keepaspectratio = true,width = 0.8\linewidth ]
    \caption{Radial position ($r$), heliographic longitude ($\phi$) and pitch angle ($\alpha$) of a 300 MeV proton for the first 5 minutes of propagation under two different scattering conditions. The left column is for a scatter-free simulation, and the right column is for a proton from a simulation with a scattering mean free path of $\lambda = 0.1$ AU. Both simulations had injection at $r_i = 20 \: \textrm{R}_\odot$. Two scattering events occur in the right hand side panels at $\sim 3.5$ and $\sim 4$ minutes.}
    \label{part_param-scatter-free}
\end{figure}

 The mirroring process as derived from our test particle code can be seen in Figure \ref{part_param-scatter-free}, where the radial distance ($r$, top panels), heliographic longitude ($\phi$, middle panels) and the pitch angle ($\alpha$, bottom panels) are displayed for the first 5 minutes of propagation for particles in two different simulations. The left panels are for a proton that propagated scatter-free and the right panels are for a proton with a scattering mean free path of $\lambda = 0.1$ AU. In both simulations each proton was injected at $r_i = 20 \: \textrm{R}_\odot$. Scattering produces an abrupt change in trajectory by reassigning the velocity vector of the particle, altering the pitch angle (at $t \sim 3.5$ minutes and $t \sim 4$ minutes in the right panels of Figure \ref{part_param-scatter-free}). The changes in longitudinal position (measured in a stationary frame not corotating with the Sun) of the proton are due to drifts that occur as the proton propagates through the IMF \citep{dal_2013}, the corotation of the system and a small curvature of the Parker Spiral. Adiabatic deceleration was negligible over the timescales depicted in Figure \ref{part_param-scatter-free}.

%Top panel: the change in radial position of the proton as it mirrors. Mirroring occurs where the radial height is at a minimum and the pitch angle is at $90^\circ$. Middle panel: The change in longitude over the first 3 minutes of the proton's propagation. The particle travels westwards as it propagates towards the sun prior to mirroring and eastwards along the Parker spiral after mirroring, where it propagates away from the sun. Bottom panel: The pitch angle of the proton over the first three minutes of its propagation. The pitch angle evolves from its initial value at injection of $\sim 127^\circ$ towards $90^\circ$ where the proton mirrors. The proton's pitch angle continues to decrease as it focused along the Parker spiral as it propagates away from the sun. The protons from this simulation were injected at $r_i = 20 R_\odot$ under scatter-free condition ($\lambda = 0.0$ AU)

\begin{figure}[ht!]
    \centering
    \includegraphics[keepaspectratio = true, width=1.0\linewidth]{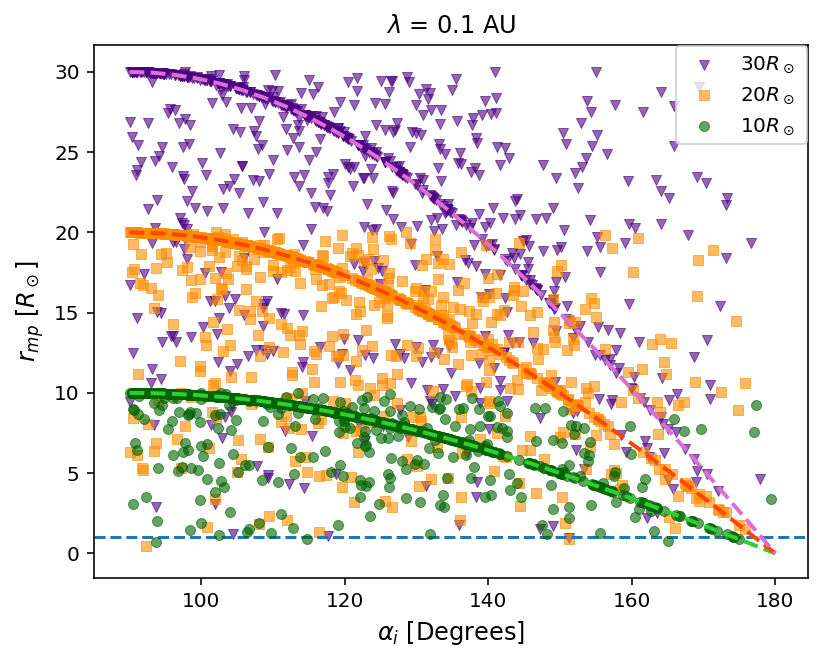}%[keepaspectratio = true, width=1.0\linewidth] [keepaspectratio = true, width=0.8\linewidth]
    \caption{Mirror point radius, $r_\mathit{mp}$, versus particle initial pitch angle, $\alpha_i$ for simulations with injections at 10 R$_\odot$ (\textit{green circles}), 20 R$_\odot$ (\textit{orange squares}) and 30 R$_\odot$ (\textit{purple triangles}) with $\lambda =$ 0.1 AU. The green, orange, and purple dashed lines show the analytical, scatter-free mirror point radius, given by Equation \ref{r_mp-B-ratio-in-text}. The horizontal blue dashed line denotes the required depth in the solar atmosphere the protons must reach to generate $\gamma$ rays (1 R$_\odot$).}
    \label{multi-rad-MP}
\end{figure}

To study the effect of scattering on the mirror point radius, we simulated the propagation of a population of 300 MeV protons with mean free path $\lambda = 0.1$ AU.
For each proton injected into the simulation the mirror point was determined by identifying the minimum radial distance the particle reached. Figure \ref{multi-rad-MP} shows the mirror point radius, $r_\mathit{mp}$ versus initial pitch angle $\alpha_i$ for the cases $r_i = $ 10, 20 and 30 R$_\odot$. For clarity, only a subset of data points were plotted for each simulation. The purple, orange and green dashed lines give the scatter-free analytical value of $r_\mathit{mp}$, according to Equation \ref{r_mp-B-ratio-in-text}. A number of data points were found to lie along these lines, corresponding to protons that did not scatter in the simulation, validating our code and methodology for deriving $r_\mathit{mp}$. The blue dashed line at $1$ R$_\odot$ displays the height in the solar atmosphere that the protons must reach to go on to generate $\gamma$ rays. Data points not along the curved dashed lines correspond to protons that experienced scattering. %(1.5\% of the 50000 calculated mirror point radii)

In Figure \ref{multi-rad-MP} there were a number of data points at or below the $1 \: \textrm{R}_\odot$ blue dashed line; these protons would be candidates to go on to produce $\gamma$ rays. It is clear from Figure \ref{multi-rad-MP} that they are a small fraction of the population. In the scatter-free case only particles in the loss cone ($\alpha_i$ close to $180^\circ$) reach the solar surface, while for the scattering case particles across the pitch angle distribution can reach it if scattered favourably.

Figure \ref{multi-rad-MP} shows that scattering allows the possibility for protons to propagate deeper into the solar atmosphere than they would have in scatter-free conditions (indicated by the points below the corresponding dashed line). 
These protons' velocity vectors have been scattered such that their pitch angles have shifted towards $180^\circ$ (i.e. more field aligned). However, there were also a number of points above the scatter-free prediction, where pitch angles were shifted close to $90^\circ$ (or the particle was reflected by the scattering event), resulting in the proton mirroring further from the solar surface than it would have under scatter-free conditions. The question of whether scattering primarily helps or hinders protons in their back-precipitation to the photosphere is addressed in Section \ref{subsec: P_vs_lambda}.

%====================================================================================================== 

\section{Precipitation from instantaneous injection at a given injection height}\label{sec:Results}%Instantaneous precipitation fraction

For a simulation in which $N$ particles were injected at height $r_i$, we define the instantaneous precipitation fraction $P$ as the percentage of the injected population that reached $r_p$,  $P$=$N_p$/$N$ $\times 100$, where $N_p$ is the number of protons that successfully precipitate.
Unless otherwise specified, we study precipitation to $r_p = 1$ R$_\odot$. In our simulations we assume the velocity distribution to be isotropic at the location of injection. Table \ref{sim-table} gives values of $N_p$ and $P$ for our simulations.

\subsection{Scatter-free precipitation fraction}
We first considered precipitation in the scatter-free case. In general, for an isotropic particle population propagating in a magnetic field increasing monotonically between $r_i$ and $r_p$, the precipitation fraction is given by

\begin{equation}
    {P_{sf} = \frac{1}{2} \left[1 - \sqrt{1 - \frac{B_{i}}{B_p}}\right] \times 100 \:\:\: (\%)},
    \label{precip_eqn-general}
\end{equation}
where $B_i$ is the magnetic field strength at the position of injection and $B_p$ is the magnetic field strength at the precipitation radius.

Close to the Sun the Parker spiral magnetic field can be approximated as a purely radial field (as given by Equation \ref{parker_field} with $\Omega = 0$). In this case for an isotropic proton population Equation \ref{precip_eqn-general} becomes

\begin{equation}
    P_\mathit{sf} = \frac{1}{2} \left[1 - \sqrt{1 - \left(\frac{r_{p}}{r_i}\right)^2}\right] \times 100 \:\:\: (\%),
    \label{radial_precip_eqn}
\end{equation}
where $r_i$ is the radius of injection. Considering $r_i = 20$ R$_\odot$ and $r_{p} = 1$ R$_\odot$ Equation \ref{radial_precip_eqn} yields $P_\mathit{sf} = 0.063$\%.

%========================================================
For a Parker spiral magnetic field (Equation \ref{parker_field}), not necessarily close to the Sun, Equation \ref{precip_eqn-general} gives
\begin{equation}
P_\mathit{psf} = \frac{1}{2}  \left(1 - \left[1 -\sqrt{\frac{r_i^{-4} + (a r_i)^{-2}}{r_p^{-4}+ (a r_p)^{-2}}}\right]^\frac{1}{2}\right) \times 100 \:\:\:  (\%).
\label{Parker_radial_precip_eqn}
\end{equation}

Considering $r_i = 20$ R$_\odot$ and $r_{p} = 1$ R$_\odot$ Equation \ref{Parker_radial_precip_eqn} also yields $P_\mathit{psf} = 0.063$\%. We carried out scatter-free simulations with $r_i = 20 \: \textrm{R}_\odot$ using our model and obtained an instantaneous precipitation fraction $P = 0.062$\% (Table \ref{sim-table}) in good agreement with the analytical value.

\subsection{Precipitation fraction in the presence of scattering}\label{subsec: P_vs_lambda}

% \begin{figure*}[ht!]
%     \centering
%     \includegraphics[width = 0.9\linewidth, height = 7cm]{1-2_5-boundary-same-panel-ri20-70.png}
%     \caption{Left panel: Instantaneous precipitation fraction ($P$) versus scattering mean free path $\lambda$ for $r_{p} = 1$ R$_\odot$, (\textit{blue points}) and $r_{p} = 2.5 R_\odot$ (\textit{red points}) from simulations over a 24 hour period that were injected at a height of $r_i = 20$ R$_\odot$, $\lambda$ ranges from $\lambda = 0.0075$ AU  ($\sim$1.6 R$_\odot$) to $\lambda = 1.0$ AU.  The solar wind speed is $v_{sw}$=500 km/s. The $r_{p} = 1$ R$_\odot$ points are fitted with the curve $P = 0.046 \left(\lambda/\text{AU}\right)^{-0.690} + 0.072$ \% and the $r_{p} = 2.5 R_\odot$ points are fitted with the curve $P = 0.344 \left(\lambda/\text{AU}\right)^{-0.596} + 0.408$ \%. Right panel: $P$ versus the $\lambda$ for injections at 70 R$_\odot$ for $r_{p} = 1$ R$_\odot$. Scattering mean free paths range from $\lambda = 0.0025$ AU to $\lambda = 0.1$ AU. The points are fit with the curve $P = (0.006/(\lambda + 0.01)) \exp \{(0.009/(\lambda + 0.01)) - ((\lambda + 0.01)/4.63 \times 10^8) \}$ \%.}
%     \label{percent_vs_mfp}
% \end{figure*}
\begin{figure*}[ht!]
    \centering
    \includegraphics[keepaspectratio = true, width=0.9\linewidth]{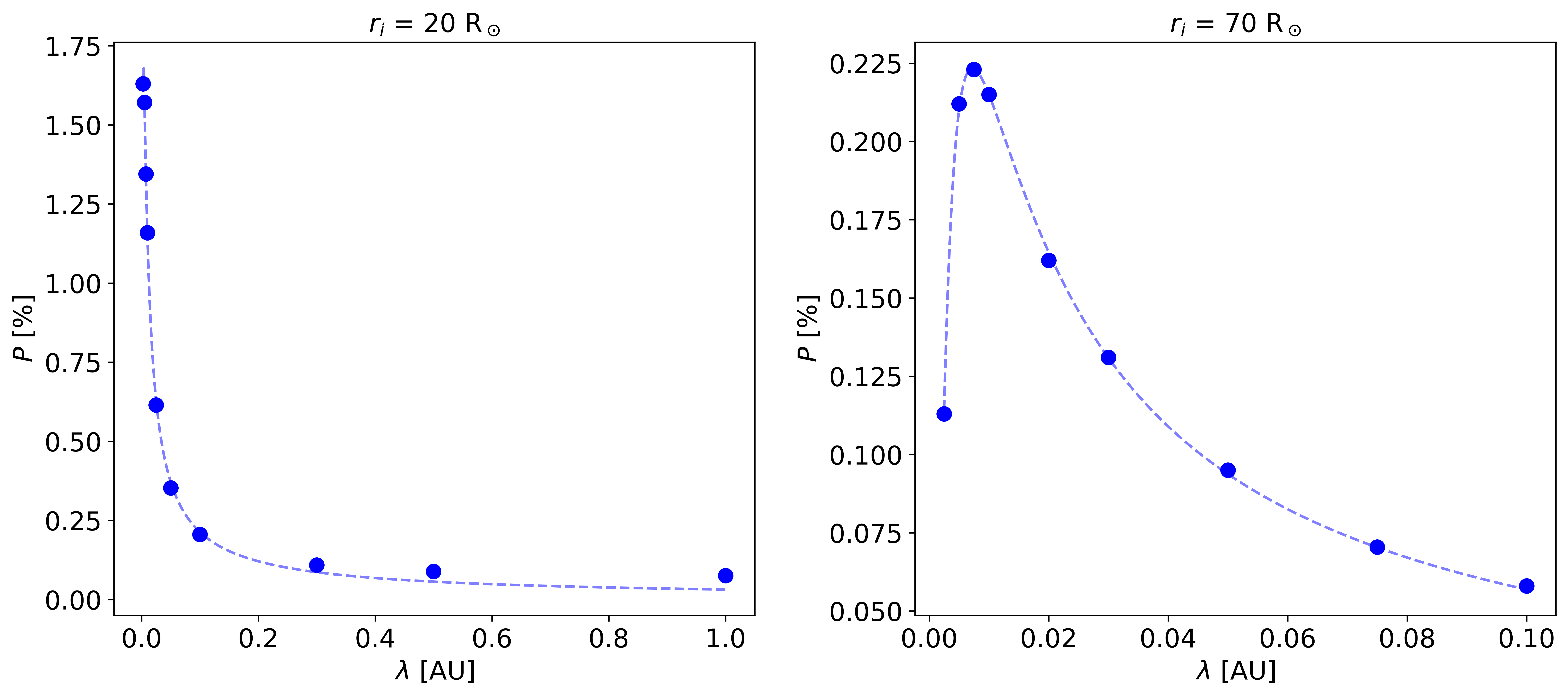}%[keepaspectratio = true, width=1.0\linewidth] [width = 0.9\linewidth, height = 7cm]
    \caption{Instantaneous precipitation fraction ($P$) versus scattering mean free path ($\lambda$). Left panel: Instantaneous precipitation fraction ($P$) versus scattering mean free path $\lambda$ from simulations with durations of 24 hours, where protons were injected at a height of $r_i = 20$ R$_\odot$, and $\lambda$ ranges from $\lambda = 0.0025$ AU to $\lambda = 1.0$ AU. The solar wind speed is $v_{sw}$= 500 km s$^{-1}$. Right panel: $P$ versus $\lambda$ for injections at 70 R$_\odot$. Scattering mean free paths range from $\lambda = 0.0025$ AU to $\lambda = 0.1$ AU. Both panels were fitted with curves of the form of Equation \ref{P(lambda)} with the left panel having constants; $a$ = 0.0318 AU, $b$ = 0.8287 and $c$ = 0.0027 AU, and the right panel having constants; $a$ = 0.0094 AU, $b$ = 0.8054 and $c$ = 0.0058 AU.
    \label{percent_vs_mfp}}
\end{figure*}% 0.0075 au ($\sim$1.6 R$_\odot$)

We carried out simulations with $N$ = 10 million protons propagating over 24 hours in a variety of scattering conditions and derived instantaneous precipitation fractions to $r_p = 1.0$ R$_\odot$.
Figure \ref{percent_vs_mfp} shows $P$ versus the scattering mean free path, $\lambda$, for injection at $r_i = 20 $ R$_\odot$ ({\it left panel}) and $r_i = 70$ R$_\odot$ ({\it right panel}).

The results show that increasing the amount of scattering does help precipitation, however, in the $r_i = 70$ R$_\odot$ case it is also evident that the efficiency of precipitation decreases at very small values of $\lambda$ after a peak value is reached.
By running a simulation at very small mean free path for $r_i = 20$ R$_\odot$ we verified that a peak in the $P$ profile at low $\lambda$ is present also in this case.

We find that a function of the form
\begin{equation} 
  P(\lambda) =  \frac{a}{\lambda^{b}} \; \exp{\left(-\frac{c}{\lambda}\right)}
    \label{P(lambda)},
\end{equation}
where $a$, $b$ and $c$ are positive constants, provides a good fit to the simulation points ({\it blue dashed lines}).

The peak in precipitation fraction results from the fact that when the scattering becomes very strong, 
outward convection with the solar wind overcomes the positive effects of enhanced scattering.
As confirmation of this interpretation, a function similar to Equation \ref{P(lambda)} can be obtained from solution of a transport equation including focussing, diffusion and convection, in the strong scattering limit \citep{Earl_1974}. %{\it add reference to Earl 1976}.
From a test-particle model point of view, the peak corresponds to  conditions that maximise the chances of particles scattering into the loss cone and remaining in it long enough to reach the precipitation radius. Any more scattering taking place ejects the particles from the loss cone too fast.

The position of the peak depends on injection height and solar wind speed. At larger $r_i$ the peak is reached at a larger $\lambda$ because scattering effects have more time to play a part.
We found from our simulations that $P$ depends weakly on the solar wind speed, increasing with decreasing $v_{sw}$, when all other parameters were kept constant.
The value of $\lambda$ at which $P$ reaches a peak decreases for decreasing $v_{sw}$.

For $r_i$= 70 R$_\odot$ (\textit{right panel}) $P$ reaches a peak value of $P \sim 0.22$\% at $\lambda \sim 0.0072$ AU. 
In the case $r_i = 20 $ R$_\odot$ (\textit{left panel}), the fit indicates that the peak in precipitation fraction would be $P \sim 1.69$\% at $\lambda$ = 0.0032 AU.

For medium/high values of $\lambda$ (low scattering) there is a tendency for protons to precipitate soon after injection, which is quantified in the final column of Table \ref{sim-table}. For the high $\lambda$ case this is especially important as the protons that precipitate early represent a very large percentage of the total number of successfully precipitating protons. We define $F_{10min}$ as the percentage of the total precipitating protons over the full simulation (24 hours), $N_p$, that precipitate in the first 10 minutes. The last column in Table \ref{sim-table} gives $F_{10min}$ for our simulations. For a simulation with $r_i$ = 20 R$_\odot$ and $\lambda$ = 0.1 AU $F_{10min} $ $\sim 69 \%$. For the same injection location, when $\lambda = 0.01$ AU, this drops dramatically to $F_{10min} $ $\sim 25 \%$. However, a greater number of protons than the $\lambda = 0.1$ AU simulation reach the solar surface over the first 10 minutes for $\lambda = 0.01$ AU because precipitation is more efficient. As expected, more turbulent magnetic fields lead to smaller $F_{10min}$ values as protons are slowed due to more frequent scattering events. Simulations with injections close to the solar surface have high $F_{10min}$ values, for instance $F_{10min} $ $\sim 96\%$ for a simulation with, $r_i = 5$ R$_\odot$, $\lambda = 0.1$ AU. This percentage decreases with increasing radial position of the injection region (see Table \ref{sim-table}).  

 In addition to modelling back-precipitation to $r_p = 1$ R$_\odot$ we also ran simulations with protons propagating to $r_p = 2.5$ R$_\odot$, corresponding to the height of the source surface, from where a model of coronal and photospheric magnetic fields would need to be used to obtain a more precise estimate of precipitation fractions to the photosphere. Precipitation fractions to the source surface were found to be about 5 times larger compared to those shown in Figure  \ref{percent_vs_mfp}.

\subsection{Dependence of $P$ on injection height} \label{subsec:rad_dependence}

\begin{figure}[h]
    \centering
    \includegraphics[keepaspectratio = true, width = 1.0\linewidth]{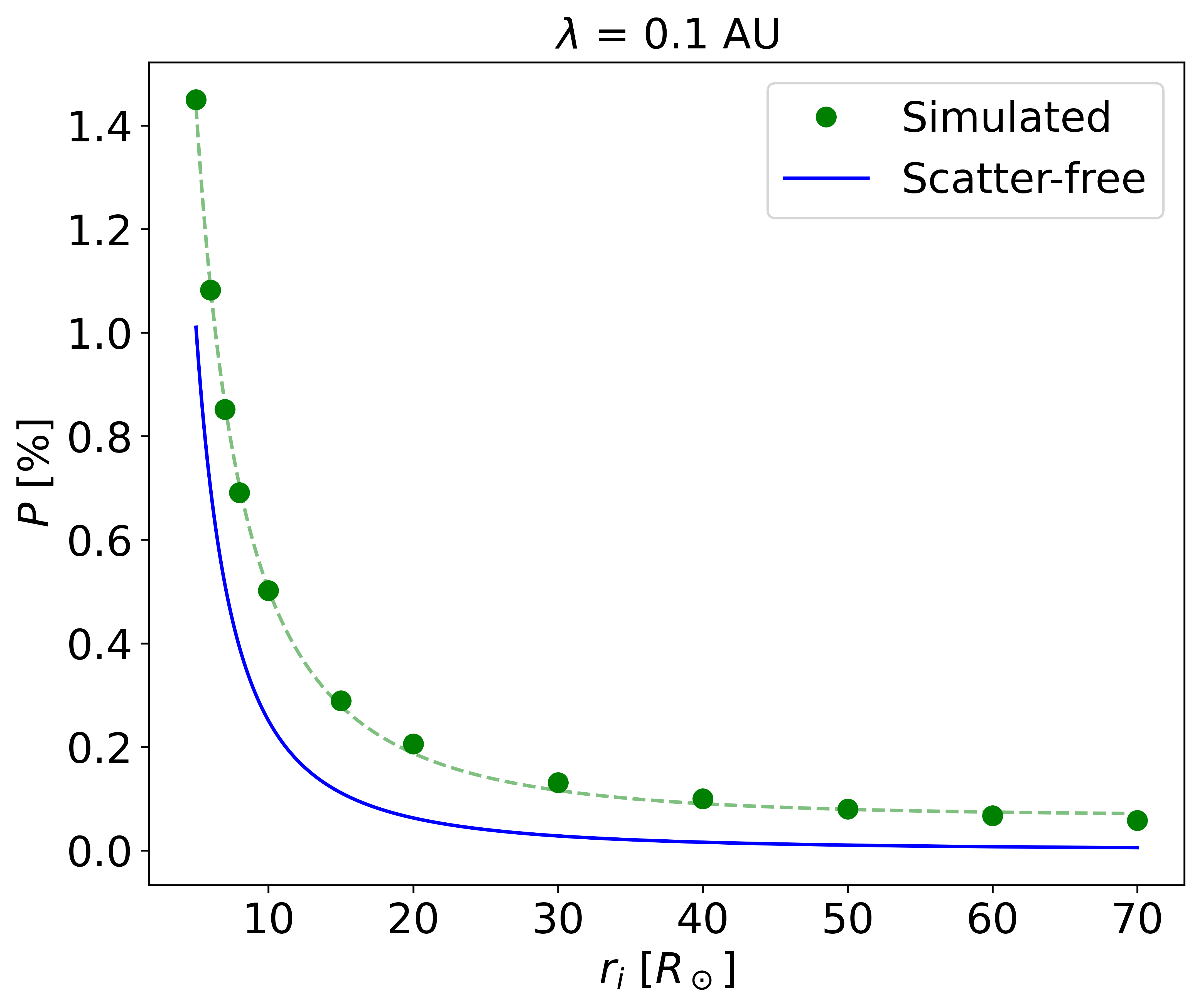}%[keepaspectratio = true, width=1.0\linewidth] [keepaspectratio = true, width = 0.8\linewidth]
    \caption{Instantaneous precipitation fraction ($P$) versus radial location of injection ($r_i$) for scattering described by a mean free path $\lambda$ = 0.1 AU. The points have been fitted with the line described by Equation \ref{P(ri)}, and the blue curve represents scatter-free precipitation according to Equation \ref{Parker_radial_precip_eqn}.
} %$P(r_i) = 41.814 \left(r_i/\text{R}_\odot\right)^{-1.777} + 0.053$
%y = 41.81439$x^{-1.77735}$ +0.05382 - including 70Rsun point
%$P(r_i) = 32.225(1-[1-\sqrt{(r_i^{-2.994} + a^{-2} r_i^{-0.052})/(r_p^{-4} + (a r_p)^{-2})}]^{1/2})$ \%
    \label{percent_vs_inject_rad_plot}
\end{figure}

Having studied the dependence of $P$ on the scattering conditions, we focussed on a specific mean free path value, $\lambda = 0.1 $ AU and investigate the radial dependence of $P$ with injection height.
Figure \ref{percent_vs_inject_rad_plot} shows $P$ versus $r_i$, %and propagated them under scattering conditions described by
characterised by a sharp decline with increasing injection height, due to the stronger magnetic mirror effect. Scattering results were compared with the 
scatter-free curve ({\it blue line}).
%{\bf Need to update this part - also tidy up the figure (eg remove fitting function text and simplify caption). Re the blue line, we note that Eq (3) is not valid at large distances from the Sun - it might be that 70 R$_{sun}$ is too far to usedit. In this case just plot the curve until eg 30 solar radii.}
The simulation points can be fit by
\begin{equation}
    %P(r_i) = 41.814 \left(\frac{r_i}{\text{R}_\odot}\right)^{-1.777} + 0.053 \:\:\: (\%)
    P(r_i) = 31.084\left(1-\left[1-\sqrt{\frac{(r_i^{-2.990} + a^{-2} r_i^{0.035})}{(r_p^{-4} + (a r_p)^{-2})}}\right]^{1/2}\right), \:\:\: (\%)
    \label{P(ri)}
\end{equation}

where $r_i$ is in solar radii. As the shock-source propagates from 5 to 70 R$_\odot$ the precipitation fraction from our simulations drops from 1.450\% to 0.058\%. This corresponds to a drop in the number of protons reaching the solar surface by a factor of $\sim25$ assuming an injection function that is constant with $r_i$.

Equation \ref{P(ri)} is a proxy for the temporal evolution of the instantaneous precipitation fraction. Depending on its speed, each CME-driven shock will cover the range of radial distances in Figure \ref{percent_vs_inject_rad_plot} over timescales that are unique to that event. Hence, the shock-height versus time curve determines the temporal evolution of the instantaneous precipitation fraction.
%, and so the temporal evolution of the instantaneous precipitation fraction for each event can be determined by considering the shock speed.

% It is important to mention that protons are able to propagate to the solar surface regardless of the level of the scattering, for the values of $\lambda$ that we considered. However, the proportion of the injected population that successfully precipitate is altered significantly by the scattering conditions, as seen in Table \ref{sim-table} and Figure \ref{percent_vs_mfp}. 

% so particles injected further from the sun are injected into a weaker magnetic field. The angle that the patricle's injection pitch angle is altered by due to the magnetic mirror effect is proportional to the ratio of the magnetic field strength at the current position to the magnetic field strength at the injection position. So 

%which is expected since the particle must propagate further, through a larger change in magnetic field strength, to produce $\gamma$ rays which subsequently leads to a smaller loss cone. This will reduce the likelihood of a proton scattering in such a way that will enter the loss cone.

%magnetic mirror effect becomes stronger for particles that travel through a larger increase in magnetic field strength. 
% Due to the inverse square law by which the magnetic field strength diminishes, protons released further away from the solar surface will have to travel a greater distance to experience the same increase in magnetic field strength. Hence, protons that are injected further from the sun are more likely to be scattered over the same change in pitch angle due to the magnetic mirror effect.

% Therefore, we expect the precipitation fractions with the inclusion of the more complex magnetic structure in the corona to be fairly similar to those determined here using a unipolar Parker spiral magnetic field. 

% the source surface and, roughly, to the radial distance where the highest-energy particles associated with Ground Level Enhancements (GLEs) are released from the shock 

\begin{table}[ht]
    \centering
    % \caption{Instantaneous Precipitation Fractions from our Simulations}
    % \centering
    \begin{tabular}{ c c c c c}%{ | c | c | c |  }%
         \hline
        {$r_i$ [R$_\odot$]} & {$\lambda$ [AU]} & {$N_p$} & { $P$ [\%]} & {$F_{10min}$ [\%]}\\ \hline
    	5 & 0.1 & 144978 & 1.450 & 95.6\\ %\hline
        6 & 0.1 & 108200 & 1.082 & 94.2\\ %\hline
        7 & 0.1 & 85170 & 0.852 & 92.4\\ %\hline
        8 & 0.1 & 69082 & 0.691 & 90.8\\ %\hline
        10 & 0.1 & 50222 & 0.502 & 87.2\\ %\hline
        15 & 0.1 & 28900 & 0.289 & 78.2\\ %\hline
        20 & 0.1 & 20614 & 0.206 & 69.2 \\ %\hline
        25 & 0.1 & 15294 & 0.153 & 59.6\\ %\hline
        30 & 0.1 & 13086 & 0.131 & 52.5\\ %\hline
        40 & 0.1 & 9950 & 0.100 & 39.9\\ %\hline
        50 & 0.1 & 8048 & 0.080 & 28.2\\ %\hline
        60 & 0.1 & 6647 & 0.066 & 19.0 \\ %\hline
        70 & 0.1 & 5798 & 0.058 & 12.7 \\ %\hline
        20 & 0.0025 & 162987 & 1.630 & 2.0 \\ %\hline
        20 & 0.0050 & 157126 & 1.571 & 10.0\\ %\hline
        20 & 0.0075 & 134501 & 1.345 & 17.8\\ %\hline
        20 & 0.01 & 115851 & 1.159 & 24.5\\ %\hline
        20 & 0.025 & 61459 & 0.614 & 45.8\\ %\hline
        20 & 0.05 & 35324 & 0.353 & 58.7\\ %\hline
        20 & 0.3 & 10891 & 0.109 & 82.7\\ %\hline
        20 & 0.5 & 8903 & 0.089 & 87.2\\ %\hline
        20 & 1.0 & 7514 & 0.075 & 93.5\\ %\hline
        20 & Scatter-Free & 6224 & 0.062 & 100.0\\ %\hline
        70 & 0.0025 & 11317 & 0.113 & 0.0 \\ %\hline
        70 & 0.0050 & 21234 & 0.212 & 0.0 \\ %\hline
        70 & 0.0075 & 22318 & 0.223 & 0.0 \\ %\hline
        70 & 0.01 & 21548 & 0.215 & 0.0 \\ %\hline
        70 & 0.02 & 16219 & 0.162 & 0.1 \\ %\hline
        70 & 0.03 & 13130 & 0.131 & 0.9 \\ %\hline
        70 & 0.05 & 9481 & 0.095 & 4.0 \\ %\hline
        70 & 0.075 & 7061 & 0.070 & 8.2 \\ %\hline
        \hline
    \end{tabular}
    
    \caption{Instantaneous precipitation fractions for our simulations. Columns are: (From left to right) The radial height of the shock at particle injection ($r_i$),  the parallel scattering mean free path ($\lambda$), the number of precipitating protons that reach $1$ R$_\odot$ over the full 24 hour simulation ($N_p$), the instantaneous precipitation fraction ($P$), and the percentage of the precipitating protons that reach the solar surface in the first 10 minutes after injection ($F_{10min}$).  All simulations injected a 300 MeV mono-energetic proton population consisting of $N$ = 10 million protons into an $8^\circ \times 8^\circ$ injection region.} 
    \label{sim-table}
\end{table}

% \begin{figure*}[ht]
%     \centering
%     \includegraphics[width=0.82\linewidth, keepaspectratio = true]{dual_hist_10min.png} %height=8cm
%     \caption{Histogram of the precipitation rate ($P_{rate}$, the number of 300 MeV protons reaching the solar surface per second) over the first 10 minutes for simulations from single injections at an injection radii of $r_i = 20$ R$_\odot$ with scattering mean free paths of 0.1 AU (left) and 0.01 AU (right). }
% \label{10min-hists}
% \end{figure*}

\section{Evaluation of emission region features}\label{subsec:long_lat_pos}
%****************************************************************************
%******** BE AWARE LAT vs LONG PLOT PLACED BEFORE THIS SECTION **************
%****************************************************************************
\begin{figure*}[ht!]
    \centering
    \includegraphics[keepaspectratio = true, width = 0.75\textwidth]{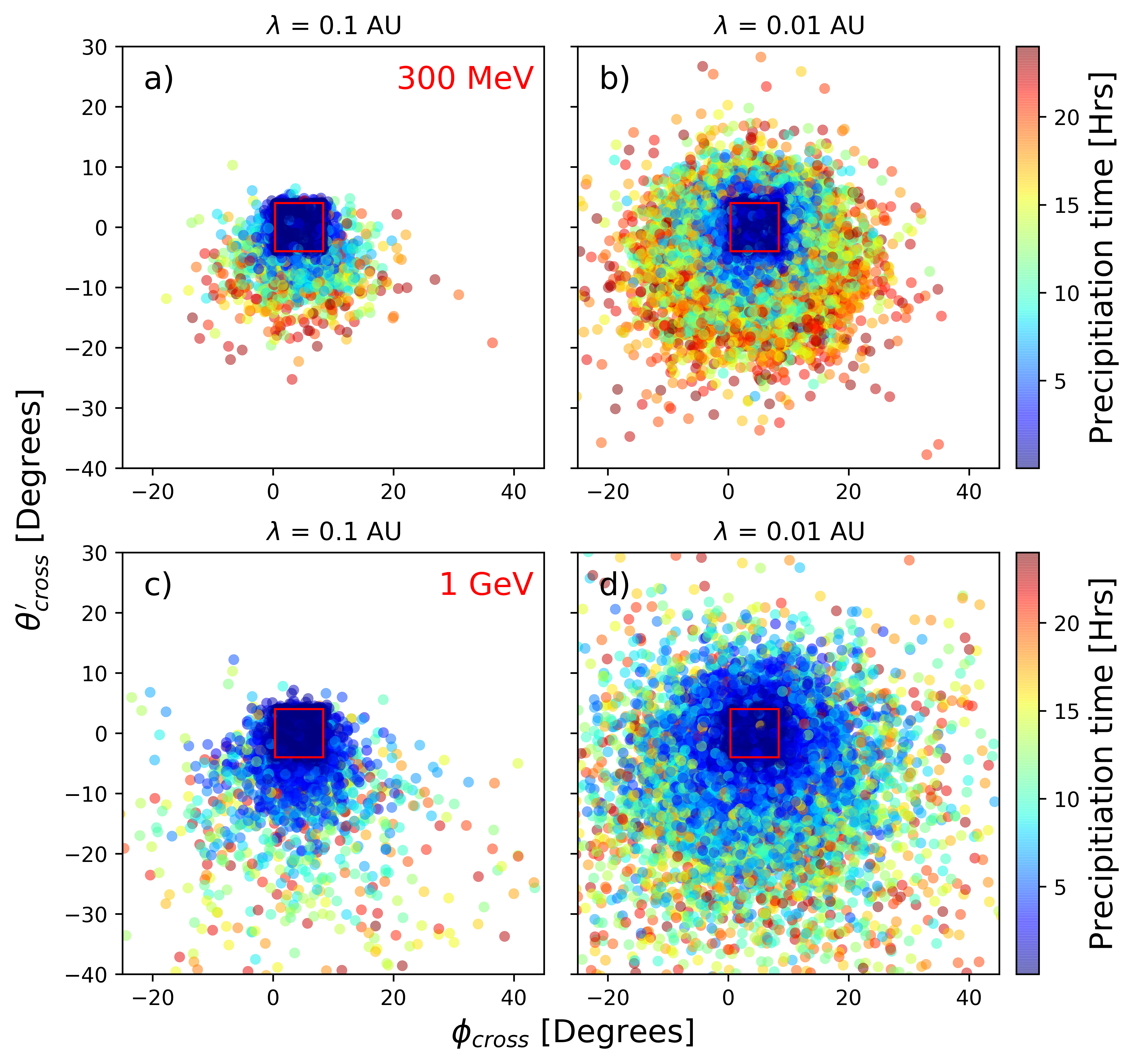}% [keepaspectratio = true, width = 0.75\textwidth] [keepaspectratio = true, width = 0.78\textwidth]
    \caption{Heliographic latitude ($\theta^\prime_{cross}$) and longitude ($\phi_{cross}$) of the locations where energetic protons  reach the solar surface for simulations with; an $8^\circ \times 8^\circ$ injection region (centred at $0^\circ$, $0^\circ$) at  $r_i = 20$ R$_\odot$ and $\lambda$= 0.1 AU (panels a and c) and 0.01 AU (panels b and d). Panels a) and b) are for a 300 MeV mono-energetic proton population, while c) and d) are for a 1 GeV population. The red box denotes the region on the solar surface that maps to the injection region. All panels are displaying -40 to 30 degrees in latitude and -25 to 45 degrees in longitude. The crossing positions of the 1 GeV simulations extend beyond these limits with particles in the $\lambda = 0.01$ AU simulation crossing 1 R$_\odot$ between -82.5 to 97.6 degrees in longitude and between -79.1 and 67.3 degrees in latitude.} %for easy comparison with panels \textit{a} and \textit{b}
  
    \label{lat-vs-long-20Rsun-0_1au-plot}
\end{figure*}

In Figure \ref{lat-vs-long-20Rsun-0_1au-plot} the locations where the protons crossed the $1$ R$_\odot$ boundary are displayed for two particle energies (300 MeV (top panels) and 1 GeV (bottom panels)) and scattering mean free paths of $\lambda = 0.1$ AU  (panels a and c) and $\lambda = 0.01$ AU  (panels b and d) in a coordinate system corotating with the Sun. The red square is the region of the photosphere directly connected to the injection region. All simulations had injections located at $r_i = 20$ R$_\odot$.
% The westward drift is associated with corotation of magnetic flux tubes over time.

In Figure \ref{lat-vs-long-20Rsun-0_1au-plot} panels a) and c) there is a systematic drift of the crossing position southwards with time. The same trend is seen in Figure \ref{lat-vs-long-20Rsun-0_1au-plot} panels b) and d). However, some crossings were also observed northwards of the emission region. The southward drift is due to gradient and curvature drifts associated with the Parker spiral magnetic field \citep{dal_2013}. For our magnetic polarity this leads to a southward drift; however, in the opposite magnetic polarity these drifts would be northwards. In addition, finite Larmor radius effects associated with scattering events produce motion of the guiding centre perpendicular to the magnetic field. The finite Larmor radius effects were more prominent in the $\lambda = 0.01$ AU panels and they were the cause of the northward displacement. We note that the finite Larmor radius effects do not have a preferential direction unlike the drifts. They were expected to be more prominent in the more magnetically turbulent simulations as the proton can shift up to 2 Larmor radii per scattering event and there were more scattering events in these simulations. 

The longitudinal and latitudinal positions of the protons reiterate that they do not propagate solely along the Parker spiral lines they were initially accelerated on. Therefore, drifts and finite Larmor radius effects are not negligible when determining if protons from the shock are responsible for the observed position of the $\gamma$ ray emission region on the solar disc, especially in the high scattering case. 
Higher-energy protons show stronger deviations from their original Parker spiral field lines, even for earlier precipitating protons, due to the increase in drifts and finite Larmor radius effects with increasing particle energy. This is clear when comparing panels c) and d) with panels a) and b) in Figure \ref{lat-vs-long-20Rsun-0_1au-plot}.

When considering injections at larger radial distances we found that, as expected, the emission region moves westwards. This occurs as the CME-driven shock propagates to larger radial distances it injects particles onto Parker spiral lines with footpoints located increasingly westwards on the photosphere.

We note that we have only considered a unipolar Parker spiral magnetic field. Positions on the solar disc are likely to be altered by the more complex coronal magnetic field structure.

%========================================================================================%
\section{Shock heights during LDGRF events} \label{sec:shock_heights} %at times of $\gamma$ ray emission

We now apply the results of our simulations to a set of eight specific LDGRF events: the subset of events from the \cite{Win2018} study with a $>$2 hour duration. These events are listed and discussed further in section \ref{subsec:indiv_event} (see Table \ref{avg_p_perc-table}).
They were associated with CMEs with plane-of-the-sky speeds ranging from 950 to 2684 km s$^{-1}$ based on the Coordinated Data Analysis Workshops (CDAW \footnote[1]{https://cdaw.gsfc.nasa.gov/CME\_list/halo/}) catalogue of observations by the Large Angle and Spectrometric Coronagraph (LASCO) on board the Solar and Heliospheric Observatory (SOHO).

 For each of the eight LDGRF events we estimate the position of the shock at the time of peak and end $\gamma$ ray emission, using CME data and a series of approximations.
We assumed that the shock height and speed coincide with those of the associated CME and that all parts of the shock propagate radially. We used the linear fit in the CDAW catalogue to determine height versus time for $r < 30$ R$_\odot$. At larger distances we used the empirical expression of \cite{Gop2001} to describe the shock's acceleration (or deceleration), $a_{sh}$, during the propagation to 1 AU:
 \begin{equation}
     a_{sh} = 2.193 \: - \: 0.0054 \: v_{sh},
     \label{shock_acceleration}
 \end{equation}
 where $a_{sh}$ is in m s$^{-2}$ and $v_{sh}$ is the shock speed in km s$^{-1}$. According to this equation shocks faster than $\sim 406$ km s$^{-1}$ decelerate. All the shocks listed in Table  \ref{avg_p_perc-table} were significantly faster than this and so decelerate during their propagation. 
To obtain the peak times of the $\gamma$ ray  emission we used data from Table 3 of \cite{Sha2018} and the corresponding plots in their Appendix C, and for the end times we used data from Table 1 of \cite{Win2018}.

%This corresponds to an average CME-driven shock speed of $\sim 2033$ km s$^{-1}$ (not including the slower shock speed from the 2012-03-07 event).

%the LDGRFs with faster CME-driven shocks tend to produce longer duration observed $\gamma$ ray emission.

 %In their statistical study of LDGRFs \cite{Win2018} have provided CME-driven shock plane-of-the-sky speeds that range, for LDGRFs with durations of 2 hours or greater, from 950 km s$^{-1}$ to  $2684$ km s$^{-1}$.  

\begin{figure}[h!]%width = 1.0\linewidth, height = 7cm
    \centering%keepaspectratio = true, scale = 0.5
    \includegraphics[width = 1.0\linewidth, keepaspectratio = true]{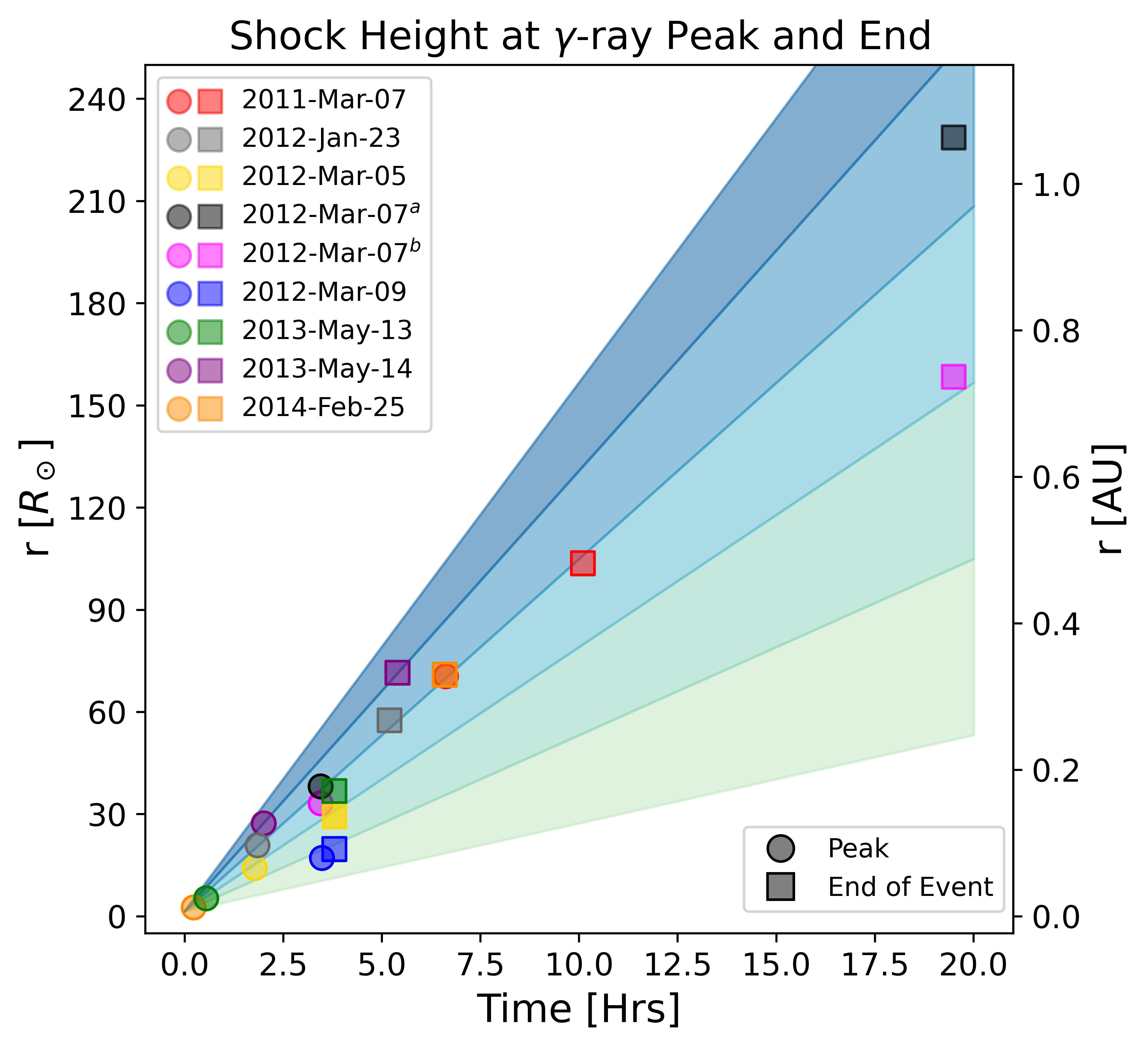}%[width = 1.0\linewidth, keepaspectratio = true] [width = 0.8\linewidth, keepaspectratio = true]
    \caption{Radial distance, $r$, from the centre of the Sun of CME-driven shocks associated with LDGRFs at the peak times (\textit{circles}) and end times (\textit{squares}) of detected $\gamma$ ray emission. Constant deceleration is assumed during propagation, as given by Equation \ref{shock_acceleration}, and initial plane-of-the-sky speeds were taken from the SOHO/LASCO CME catalogue. Each shaded wedge spans 500 km s$^{-1}$ increments in radial shock speed between 500 and 3000 km s$^{-1}$. For the 2012 March 7 event we show two sets of radial positions for two different CME-driven shocks that day: 2012-Mar-07$^a$ is the earlier, faster CME with a speed of 2684 km s$^{-1}$ and 2012-Mar-07$^b$ is the second, slower CME with a speed 1825 km s$^{-1}$.}
    \label{shock_position}
\end{figure}

For the eight LDGRF events the shock positions at the peak and end times of the $\gamma$ ray emission were plotted in Figure \ref{shock_position} as circles and squares respectively. Here the shaded wedges span 500 km s$^{-1}$ increments in constant shock speed from 500 to 3000 km s$^{-1}$. 
For the 2012 March 7 event, which was associated with two fast CMEs erupting in rapid succession, data points for both (with speeds of 2684 km s$^{-1}$ and 1825 km s$^{-1}$, respectively) were plotted. While the absence of interplanetary type III radio emissions during the second, much slower CME suggests that it was unlikely associated with the SEP event at 1 AU (see \cite{rich_2014}), a direct contribution to particle acceleration cannot be ruled out, so both CMEs were considered. 

Figure \ref{shock_position} shows that if the CME-driven shock was the source of the $\gamma$ ray emission in the LDGRF events, back-precipitation from large distances needs to have taken place.  The median position of the shock at the end time is $\sim 71$ R$_\odot$ (including both shocks that originated on 2012 March 7).
%mean position of the shock at the end time is $\sim 86$ R$_\odot$.
For the 2012 March 7 event the two associated shocks were located at $\sim 242$ R$_\odot$ and $\sim 155$ R$_\odot$ at the end time of the  $\gamma$ ray emission, which lasted for 19.5 hours.

%which had two shocks associated with it, one with a shock speed of $\sim 1825$ km s$^{-1}$ (plotted in Figure \ref{shock_position}) and another with a speed of $2864$ km s$^{-1}$ (not plotted). This event lasted $\sim$ 19.5 hours and so at the end of the event these shocks would be located $\sim 155$ R$_\odot$ and $\sim 242 $R$_\odot$ from their formation position respectively.

%We note that in Figure \ref{shock_position} the shock positions at peak time differ greatly, and likely relies on other factors affecting the acceleration efficiency, such as the shock speed and geometry, the coronal magnetic field strength, the presence of seed particle populations and pre-existing turbulence.

%We will show in section \ref{subsec:rad_dependence} that these fast shocks associated with long duration $\gamma$ ray emission experience a dramatic drop in precipitation efficiency as they propagate large distances from the solar surface. 

% \begin{itemize}
%     \item Figure \ref{shock_position} contains only events in Table 1 of \cite{Win2018} that $> 2$hrs duration and propagated the furthest
% \end{itemize}

\section{Total precipitation fraction}\label{subsec:indiv_event}
%===============================================================
%===========  avg Percent from relation stuff ==================
%===============================================================

 In this Section we combine our simulation results for the instantaneous precipitation fraction as a function of $r_i$ (as summarised in Figure \ref{percent_vs_inject_rad_plot}) with the information on shock height versus time for LDGRF events to obtain an upper limit estimate, $\overline{P}$, of the total precipitation fraction in the events, within the CME shock scenario. We calculate $\overline{P}$ as 
\begin{equation}
    \overline P = \frac{\displaystyle \int_{r_{ini}}^{r_{fin}} N(r_i) P(r_i) \: dr_i}{\displaystyle \int_{r_{ini}}^{r_{fin}} N(r_i) \: dr_i} \:\:\:\: (\times 100 \:\%), 
    \label{tot_precip-w-inj-eqn}
\end{equation}
%\displaystyle - makes larger symbols
where $P(r_i)$ describes the radial evolution of the instantaneous precipitation fraction (Equation \ref{P(ri)}), $N(r_i)$ the number of injected particles as a function of radial distance (injection function), $r_{ini}$  the radial position of the shock when particle acceleration begins, and $r_{fin}$ the position when particle acceleration ends. Hence, $\overline{P}$ is the ratio of the total number of precipitating and injected protons over the event. 
%Because of their different speeds, CME-driven shocks travel from $r = 5$ to $r = 70 \: \textrm{R}_\odot$ (the radial distance displayed in Figure \ref{percent_vs_inject_rad_plot}) over timescales that are specific to each event.

For  $P(r_i)$ we make use of the fit of the curve in Figure \ref{percent_vs_inject_rad_plot} (Equation \ref{P(ri)}), corresponding to scattering conditions described by $\lambda = 0.1$ AU, (i.e. strong scattering).
We used a simple model for the injection function given by
\begin{equation}
    % N(r_i) = \frac{1}{(r_i-r_{ini})} \exp \left(\frac{-A}{(r_i-r_{ini})} - \frac{(r_i-r_{ini})}{B}\right) % 1 should be 1 R$_\odot$
    N(r_i) = \frac{A}{(r_i - r_{ini})^B} \exp{\left(-\frac{C}{(r_i - r_{ini})}\right)},
    \label{Reid-Axford-like-inj-eqn}
\end{equation}
where $A$, $B$ and $C$ are positive constants. This functional form describes a fast rise phase to peak injection, then a turnover and subsequent slow decay, describing the fact that energetic particles are more efficiently accelerated when the shock is closer the Sun. In particular, in the scatter free hypothesis and the CME shock acceleration scenario the highest-energy particles associated with Ground Level Enhancements are thought to typically be released at shock heights of 2-4 R$_\odot$ \citep{Reames_2009,Gopal_2013}.

%For a constant injection function this simplifies to,

% \begin{equation}
%     \overline P = \frac{1}{r_{fin} - r_{ini}}\left(\int_{r_{ini}}^{r_{fin}} P(r_i) \: dr_i\right) \:\:\:\: (\times 100 \:\%)
%     \label{avg_precip}
% \end{equation}

\begin{table*}[ht]
    \centering
    % \caption{Total Precipitation Fractions for Individual Events}
    \begin{tabular}{ c c c c c c}
         \hline
        {Date} & {$v_{sh}$ [km $s^{-1}$]} & {C2 Time [UT]} & {D [Hrs]} & {$r_{fin}$ [$R_\odot$]} & {$\overline P \: (r_{ini}=1.20 R_\odot$) [\%]} \\ \hline
    % 	2012-01-23 & 2175 & 5.2 & 57.525 & 0.745 \\ %\hline
    %     2012-03-07$^a$ & 2684 & 19.5 & 228.596 & 0.235 \\ %\hline
    %     2012-03-07$^b$ & 1825 & 19.5 & 158.378 & 0.312 \\ %\hline
    %     2014-02-25 & 2147 & 6.6 & 70.915 & 0.618 \\ %\hline
    %     % Below are LDGRFs not common to de Nolfo and Share AND Winter
    %     % 2011-03-07 & 2125 & 10.1 & 103.627 & 0.444 \\ %\hline
    %     % 2012-03-05 & 1531 & 3.6 & 29.205 & 1.388 \\ %\hline
    %     % 2012-03-09 & 950 & 3.8 & 19.708 & 2.003 \\ %\hline
    %     % 2013-05-13 & 1850 & 3.8 & 36.744 & 1.122 \\ %\hline
    %     % 2013-05-14 & 2625 & 5.4 & 71.496 & 0.613 \\ %\hline
        
        2011-Mar-07 & 2125 & 20:00:05 & 10.1 & 103.6 & 0.603  \\ %\hline
        2012-Jan-23 & 2175 & 04:00:05 & 5.2 & 57.5 & 0.667   \\ %\hline
        2012-Mar-05 & 1531 & 04:00:05 & 3.6 & 29.2 & 0.800 \\ %\hline
        2012-Mar-07$^a$ & 2684 & 00:24:06 & 19.5 & 228.6 & 0.555 \\ %\hline
        2012-Mar-07$^b$ & 1825 & 01:30:24 & 19.5 & 158.4 & 0.574 \\ %\hline
        2012-Mar-09 & 950 & 04:26:09 & 3.8 & 19.7  & 0.932 \\ %\hline
        2013-May-13 & 1850 & 16:07:55 & 3.8 & 36.7 & 0.745   \\ %\hline
        2013-May-14 & 2625 & 01:25:51 & 5.4 & 71.5 & 0.640     \\ %\hline
        2014-Feb-25 & 2147 & 01:25:50 & 6.6 & 70.9 & 0.640    \\ %\hline
       
       %& {$\overline P (r_{ini}=1.43 R_\odot$) [\%] %or 1.43 R$_\odot$ (mean) 
       % & 0.629
       % & 0.683
       % & 0.798
       % & 0.589
       % & 0.604
       % & 0.914
       % & 0.750
       % & 0.659
       % & 0.660
        
        \hline

    \\
  
    \end{tabular}

    % \tablefoot
    \caption{Total precipitation fractions over the full duration of the eight LDGRF events plotted in Figure \ref{shock_position}. Columns are: (From left to right) The date of the LDGRF event, the speed of the shock associated with the CME ($v_{sh}$), the time of CME first appearance in the LASCO C2 field of view; obtained from the SOHO/LASCO CME catalogue, the duration of the detected $\gamma$ ray emission from the \textit{Fermi} LAT instrument (D); obtained from Table 1 of \cite{Win2018}, the position of the shock at the end time of detected $\gamma$ ray emission of the event ($r_{fin}$); determined using Equation \ref{shock_acceleration}, and the total precipitation fraction ($\overline{P}$) over the propagation of the shock from $r_{ini}$ (1.20 R$_\odot$, the median shock formation height as determined by \cite{Gop_2013}) to $r_{fin}$; determined using Equation \ref{tot_precip-w-inj-eqn}. The total precipitation fraction considers the fact that the instantaneous precipitation fraction decreases with the radial dependence as shown in Figure \ref{percent_vs_inject_rad_plot} and assumes that the scattering can be described by $\lambda = 0.1$ AU. There were two CMEs associated with the 2012 March 7 event, consequently 2 shocks were formed. CME \textit{a} entered the LASCO C2 field of view at 00:24:06 UT and was associated with a X5.4 flare; the second CME, \textit{b}, entered the LASCO C2 field of view at 01:30:24 UT and was linked to a X1.3 flare.} 
    \label{avg_p_perc-table}
   
\end{table*}

To calculate $\overline P$ we need to specify $r_{ini}$ and $r_{fin}$ in Equation \ref{tot_precip-w-inj-eqn}.
Conservatively, we assume that particle acceleration begins at the radial position of the formation of the shock. \cite{Gop_2013} determined the median shock formation height of 1.20 R$_\odot$, which we considered as a possible estimate for $r_{ini}$. We estimate $r_{fin}$ by calculating the shock position at the time when the $\gamma$ ray emission ends (based on the observed durations). The values of $r_{fin}$ for the LDGRF events are shown in column 4 in Table \ref{avg_p_perc-table} and plotted in Figure \ref{shock_position}.  %The duration of detected $\gamma$ ray emission is displayed in the fourth column.

\subsection{Influence of injection function on precipitation}

\begin{figure*}[ht!]
    \centering
    \includegraphics[keepaspectratio = true, width = 0.8\textwidth]{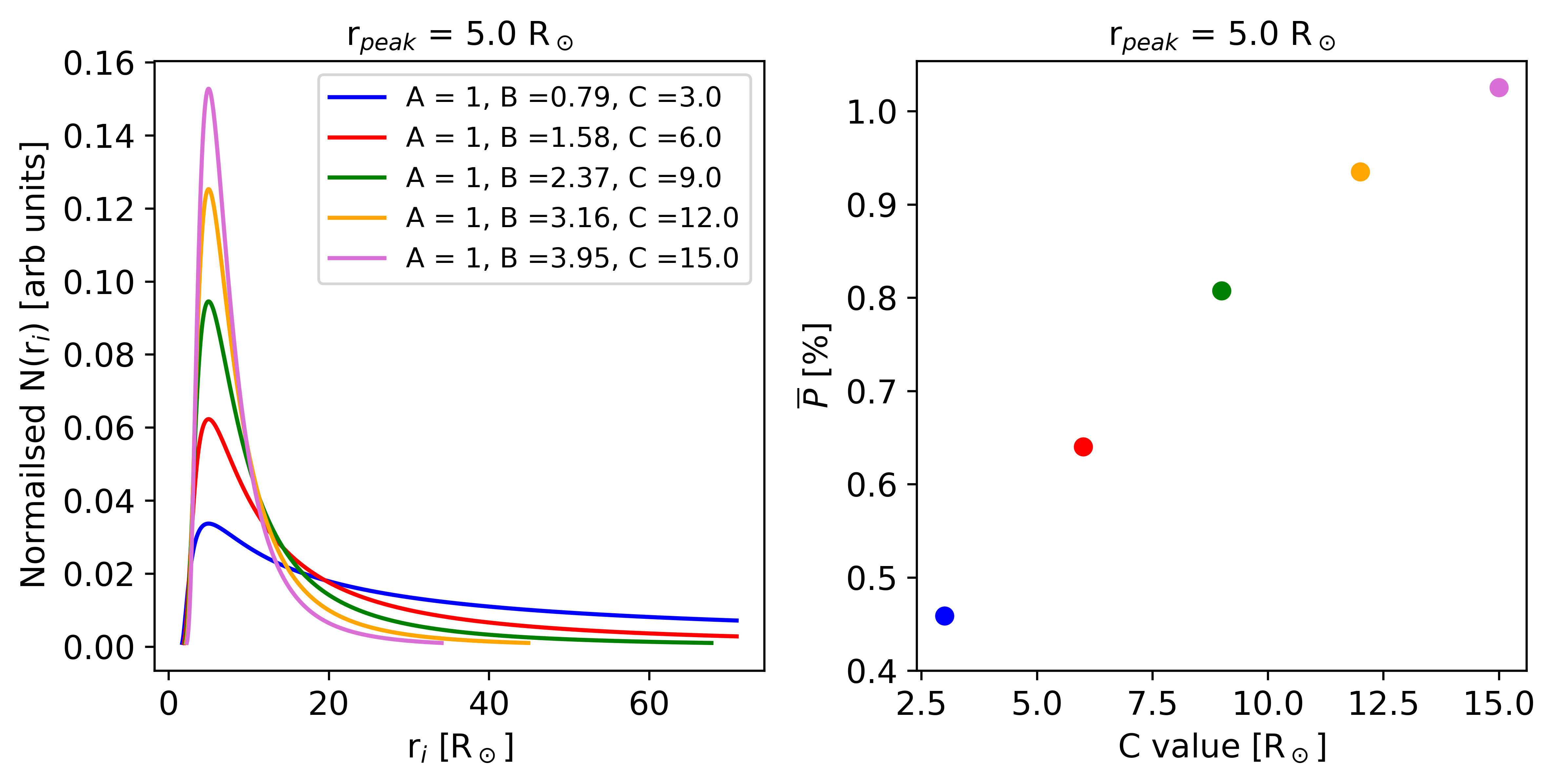}%[keepaspectratio = true, width = 0.8\textwidth] [keepaspectratio = true, width = 1.0\textwidth]
    \caption{Normalised injection functions ({\it left panel}) with peak injection at 5.0 R$_\odot$ and differing decay rates versus the radial position of the shock, $r_i$. Only values of $N(r_i) > 0.001$ were plotted for clarity. The corresponding total precipitation fraction ($\overline P$, {\it right panel}) versus the value of the constant $C$. Each coloured point in the right panel represents $\overline P$ for the same coloured curve in the left panel, calculated using $r_{ini} = 1.20$ and $r_{fin} = 70.91$ R$_\odot$, for $\lambda = 0.1$ AU.} %$86.24$ R$_\odot$ (the average end time position from the shocks in Figure \ref{shock_position})
    \label{injection_func-dual-panel-plot}
\end{figure*}

Figure \ref{injection_func-dual-panel-plot} shows the effect of injection function shape on $\overline P$. Here $N(r_i)$ for different values of $A$, $B$ and $C$ and a fixed radial position of peak injection, $r_{peak} = $ 5.0 R$_\odot$ are shown. The injection functions are normalised so that the total number of injected particles was the same for different curves. The right panel of Figure \ref{injection_func-dual-panel-plot} displays the corresponding total precipitation fractions, determined using Equation \ref{tot_precip-w-inj-eqn}, where each coloured point represents the $\overline P$ of the same coloured curve in the left panel.
Here we used $r_{ini} = 1.20$ R$_\odot$ and $r_{fin} = 70.91$ R$_\odot$, the latter being the median end time position of the shocks of Table \ref{avg_p_perc-table}.

We note that for a delta function injection the total precipitation fraction is the same as the instantaneous precipitation fraction given by Equation \ref{P(ri)}. It is clear from Figure \ref{percent_vs_inject_rad_plot} that the further from the Sun particles are injected the harder it is for them to back-precipitate. Therefore, the more sunwards the injection function is skewed the larger the total precipitation fraction is (i.e. peak injection occurring closer to the Sun leads to higher $\overline P$ and extended injections reduce $\overline P$). This is identifiable by comparing the blue and purple injection functions and their corresponding $\overline P$ in Figure \ref{injection_func-dual-panel-plot}. The blue and purple injection functions in the left panel have total precipitation fractions of 0.459\% and 1.025\%, respectively. The extended decay phase of the blue curve skews the injection further from the Sun where the efficiency of the back-precipitation is low, resulting in the lower total precipitation fraction.

% As can be seen in Figure \ref{injection_func-dual-panel-plot}, injection functions that inject a larger proportion of the energetic proton population far from the solar surface result in a smaller $\overline P$ than those that inject a large proportion of protons closer to the solar surface. 

% Even though the position of peak injection remains the same the extended decay phase of the blue curve leads to a decrease in $\overline P$ of ., \textbf{which is equal to $\sim 1.5$ times the $\overline P$ for the 2014-02-25 event (which has a shock that covers approximately the same radial distance), considering $N(r_i) = $ \textit{constant}.

% substantial compared with the majority of $\overline P$ values for a constant injection function, like those considered for the LDGRFs in Table \ref{avg_p_perc-table}.

% Similarly, injections that begin closer to the solar surface ($r_{ini}$ closer to $1.0 \: \textrm{R}_\odot$) will yield higher total precipitation fractions as back-precipitation is more efficient (i.e. maximum precipitation efficiency occurs at $r_{ini}$ for the injected particles).  % I'm not sure about this bit - needs rephrasing at least!

Figure \ref{injection_func-dual-panel-plot} shows that
for large total precipitation fractions a prompt injection close to the solar surface is required. However, this will result in shorter durations of $\gamma$ ray emission. Conversely, injections extended over large radial distances will provide longer durations of $\gamma$ ray emission at the expense of the total precipitation fraction. Therefore, large precipitation fractions and long durations of $\gamma$ ray emission cannot be reconciled using the CME-driven shock acceleration scenario.

\subsection{Total precipitation fraction for eight LDGRF events}

We calculated $\overline P$  from Equation \ref{tot_precip-w-inj-eqn} using an injection described by Equation \ref{Reid-Axford-like-inj-eqn} with $A = 1.0$ R$_\odot$, $B = 1.58$  and $C = 6.0$ R$_\odot$, corresponding to the red curve in Figure \ref{injection_func-dual-panel-plot}. The injection was normalised such that the same number of particles were considered for each injection.
The values for $\overline P$ we obtained are displayed in the final column of Table \ref{avg_p_perc-table}, corresponding to $r_{ini}=1.20$ R$_\odot$, for the eight LDGRF events.%and 1.43 R$_\odot$
 These values are small, ranging from $\overline P \sim 0.56\%$ to $\sim 0.93\%$, with the smallest $\overline P$ values associated with the events with the fastest shocks or longest durations. As indicated by Figure \ref{injection_func-dual-panel-plot} if the injection was more prompt (like the purple curve in Figure \ref{injection_func-dual-panel-plot}) then larger $\overline P$ values are possible. Similarly, if the injection was more sunwards-skewed the $\overline P$ values would increase further. We considered the effect of an injection with $r_{peak} = 3.0$ R$_\odot$, keeping all parameters the same, and $\overline P$ increased to $\sim 2$ \%.

%\textbf{Comparing the two columns one can see that when $r_{ini}$ is closer to the solar surface $\overline P$ is smaller.This effect is caused by the normalisation we use, the rise to peak precipitation is slower and the normalised injection has a lower peak and more extended decay phase, resulting in a less sunwards-skewed particle injection than the $r_{ini}=1.43$ R$_\odot$ and therefore smaller total precipitation fraction.   }

% It can been seen here that the total precipitation fractions are extremely small, ranging between \textbf{0.558\% and 0.933\%}.

% It should be noted that the height at which the shock begins to inject protons ($r_{ini}$) has a large effect on the overall value of $\overline{P}$ as the precipitation fraction increases dramatically with injection heights closer to the solar surface. For example if the shock in the 2012-01-23 event formed at 2.0 R$_\odot$, as estimated by \cite{Jos_2013}, instead of $r_{ini} = 1.43$ R$_\odot$, $\overline{P}$ would drop \textbf{from 0.431\% to 0.349\%}. Similarly, \cite{Gop_2013} computed a $1.93$ R$_\odot$ shock formation height for the 2011-03-07 event, which would correspond to a decrease in its total precipitation fraction from \textbf{0.266\% to 0.225\%}. We note that we are still operating under the assumption that particle acceleration begins at the height the shock forms. However, it takes time to accelerate protons to the $>$300 MeV energies that are required to produce $\gamma$ rays and so these $\overline P$ values are overestimates.    

The total precipitation fraction of the event, $\overline P$, becomes smaller if acceleration at the CME shock continues over larger distances, as can be seen from Table \ref{avg_p_perc-table} considering the events with the largest values of $r_{fin}$. Typically these are the events with very long durations, very fast shocks or both. Looking at the two shocks associated with the 2012 March 7 event one can see that having a slower shock over the same duration will lead to an increased $\overline P$. We note that if the proton acceleration occurs at the flanks of the shock this would increase the $\overline P$ as they remain closer to the solar surface than the shock nose. However, higher-energy particles are believed to be more efficiently accelerated over a small shock region around the nose \citep{Zan_2006,Hu_2017}.

% However, where along the shock front the majority of the high-energy protons are accelerated is not well understood \citep{Kong_2019}. The acceleration timescales are thought to be shorter at the flanks during periods of quasi-perpendicular shock geometry, but acceleration to high energies is more efficient \citep{Gia_2005}. Therefore, faster particle acceleration closer to solar surface is favoured.

\begin{figure*}[ht!]
    \centering
       \includegraphics[width=0.75\linewidth, keepaspectratio = true]{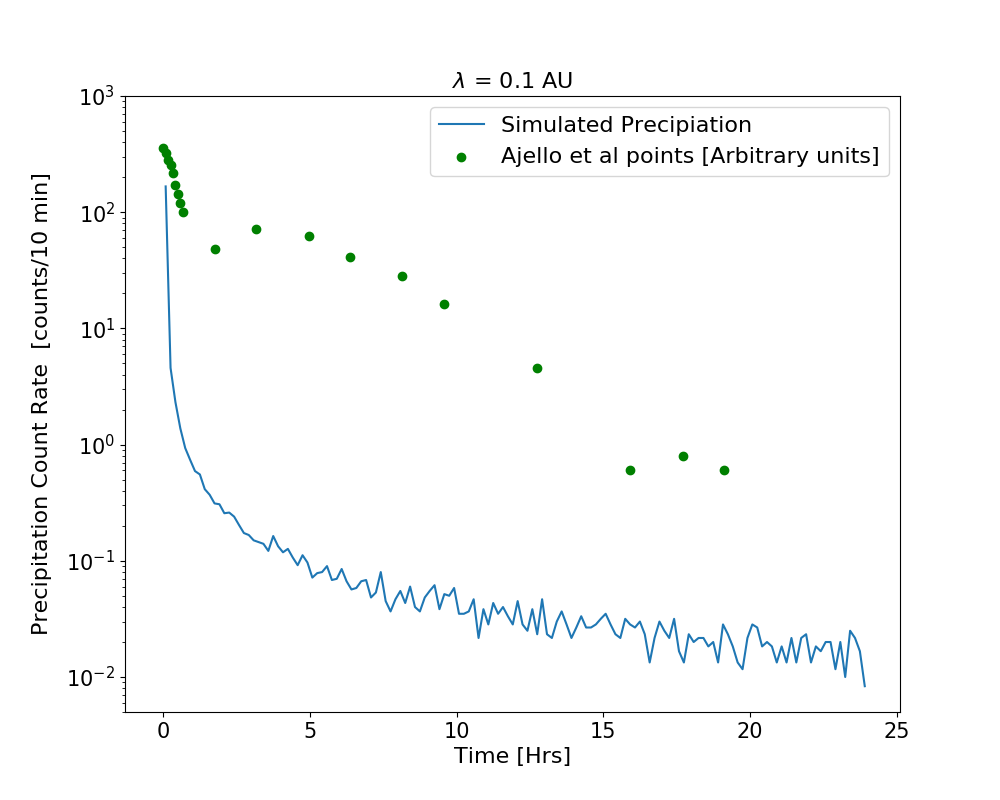}%[width=0.65\linewidth, keepaspectratio = true][width=0.95\linewidth, keepaspectratio = true]
       \caption{Precipitation count rate versus time from a simulation with injection from a moving shock-like source with the same speed as the fastest 2012 March 7 shock ({\it blue line}), at 10 minute resolution. The radial profile of  particle injection is given by Equation \ref{Reid-Axford-like-inj-eqn} with $A = 1.0$ R$_\odot$, $B = 1.58$ and $C = 6.0 $ R$_\odot$. Green data points show the time evolution of the $\gamma$ ray emission, from \cite{Ajello_2014}.}
\label{10min-hists}
\end{figure*}

%The values of $\overline P$ in Table \ref{avg_p_perc-table} are overestimates of the true precipitation fractions during the LDGRF events. We consider a number of simplifying assumptions which all contribute to larger total precipitation fractions. These assumptions are: that particle acceleration at the shock begins at the height in the solar atmosphere where the shock forms, that the LASCO CME speeds are the true radial shock speeds (they are faster closer to the solar surface as these are averaged over the LASCO field of view), we only consider a Parker spiral magnetic field geometry; whereas the highly inhomogeneous coronal and photospheric magnetic fields will result in stronger mirroring and we consider simulation times (24 hours) which are greater than the duration of detected $\gamma$ ray emission for any of the LDGRF events we have considered. Therefore, these total precipitation fractions are a best-case scenario, any further improvements on our assumptions will result in reduced total precipitation fractions.

% \subsection{Non-uniform particle injection}\label{sec:non-uni_part_inj}

\section{Time profiles of proton back-precipitation} \label{sec:Timeprofiles}

% \begin{figure*}[ht]
%     \centering
%     \includegraphics[width=0.82\linewidth, keepaspectratio = true]{iso_dual_hist_10min.png} %height=8cm
%     \caption{Histogram of the precipitation rate ($P_{rate}$, the number of 300 MeV protons reaching the solar surface per second) over the first 10 minutes for simulations from injections at an injection radius of $r_i = 20$ R$_\odot$ with scattering mean free paths of 0.1 AU (left) and 0.01 AU (right). }
% \label{10min-hists}
% \end{figure*}

%The precipitation time profile of a particle injection that is shown as the red curve in Figure \ref{injection_func-dual-panel-plot}. The main panel shows the time profile of precipitation over the 24 hour simulation and has a resolution of 10 minutes. The inset shows the time profile of precipitation over the first 10 minutes, where 90.6\% of the precipitation takes place, with a resolution of 1 second.
%(median final shock position in Table \ref{avg_p_perc-table}) 

The test particle simulations described in Section \ref{sec:Results} have considered injection at a single radial position, rather than a moving shock.
To derive information about how precipitation evolves over time, we also carried out a test-particle simulation with moving shock-like injection \citep{Hutch_2021}.
We considered a shock with radial speed of 2684 km s$^{-1}$ at 1.2 R$_\odot$ that injects particles over the radial distance range 1.2 - 228.6  R$_\odot$ and follows the injection function given in Equation \ref{Reid-Axford-like-inj-eqn} with $A = 1.0$ R$_\odot$, $B = 1.58$ and $C = 6.0 $ R$_\odot$. This injection approximates the injection for the fastest shock in the 2012 March 7 event, with the acceleration of energetic protons occurring over 19.5 hours. We assume constant shock deceleration as described in Section \ref{sec:shock_heights}.

  The time profile of precipitation can be seen in Figure \ref{10min-hists}.
 % The inset in the Figure displays  the first 10 minutes (where 90.5\% of the total precipitation takes place) with a resolution of 1 second.
  Data points from Fermi/LAT showing the time evolution of the $\gamma$ ray emission are also shown.
 It can be seen that even for an extended injection, in the simulations the majority of back-precipitation takes place early. The rapid decay in precipitation highlights the significant challenge that magnetic mirroring poses to back-precipitating protons. 

As can be seen from Figure \ref{10min-hists}, the precipitation drops by three orders of magnitude within $\sim$5 hours, even though injection continues for many hours beyond this. According to the $\gamma$ ray profile for the 2012-03-07 event \citep{Ajello_2014}, the detected $\gamma$ ray emission drops by only $\sim$2 orders of magnitude over the 19.5 hour duration of the LDGRF event. This injection considers the fastest shock associated with the 2012-03-07 LDGRF event. We also considered the contribution from the slower second shock (not displayed in Figure \ref{10min-hists}), where we find that it decays much faster than the decay of the observed $\gamma$ ray profile. %This decay cannot explain the slow decay in the $\gamma$ ray profile and the energetic proton acceleration during the event has been attributed to the faster shock \citep{rich_2014}.   

\section{Discussion and conclusions}\label{sec:Disc&conc}

 Energetic ($>$300 MeV) proton back-propagation from CME heights down to the solar surface is strongly impeded by magnetic mirroring. In this paper we investigated whether scattering associated with turbulence may aid back-precipitation to the levels required to explain LDGRFs via the CME shock scenario, by ensuring that more particles enter the loss cone.  We investigated the problem extensively using 3D test particle simulations with varying levels of scattering.

Particles accelerated at a CME-driven shock may back-precipitate via a number of different routes. There might be the possibility of almost scatter-free trajectories or propagation might be strongly influenced by the scattering behind the shock. In some cases back-precipitation from the flanks may be involved.
 We found that compared to the scatter-free case, scattering does enhance particle precipitation. For example for injection at $r_i = 20$ R$_\odot$ it increases the instantaneous precipitation fraction $P$ from 0.06\% (scatter-free) to 0.21\% for a mean free path $\lambda = 0.1$ AU.
 Increasing the level of scattering further improves the precipitation fraction to, for example $\sim P$=1.63 \% when $\lambda = 0.0025 $ AU.
 There is, however, a limit to this increase because when the scattering mean free path becomes very small outward convection with the solar wind becomes very efficient and particles can no longer back-precipitate (as shown in Figure \ref{percent_vs_mfp}). The value of the mean free path at which this effect becomes significant is dependent on injection height and solar wind speed. In our simulations the convection effect becomes important for mean free paths below 0.0072 AU for $r_i = 70$ R$_\odot$ and below 0.0032 AU for $r_i = 20$ R$_\odot$.
 
%In the analysis of the role of scattering on back-precipitation of high energy protons, it is difficult to constrain which value of the mean free path is most appropriate. 
%In the analysis of Section \ref{sec:Results} we considered a range of values of $\lambda$ from 0.0025 to 1 AU.

Some studies in the literature have assumed very strong scattering conditions behind the shock, for shock heights up to $\sim 10$ R$_\odot$. For example, \cite{Afa2018} used a model where $\lambda$ increases from 0 at the shock front up to a maximum value $\lambda_0$ at the Sun, where $\lambda_0$ varies in the range $0.16 \leq \lambda_0 \leq 3.2$ R$_\odot$ ($\sim 0.0007 \leq \lambda_0 \leq \sim 0.015$ AU) across different simulations. \cite{Jin_2018}  suggest that a mean free path of the order of 1 R$_\odot$ ($\sim 0.0047$ AU) is sufficient to overcome the strong magnetic mirroring that occurs close to the Sun. While this may be the case very close to the Sun, our results show that when the shock is further out very small mean free paths impede back-propagation via the outward convection effect. When shock locations further from the Sun are considered very low mean free path values are probably unrealistic. 
If the scattering were very strong all the way from the shock to the corona, many hours after CME liftoff, one would expect to observe a long lasting increase in SEP fluxes after the passage of the CME driven shock at a near-Sun spacecraft. To our knowledge this has never been seen, for instance in data from the Helios 1 and 2 spacecraft (e.g. \cite{kallenrode_1993}). It is hoped that new data from Parker Solar Probe and Solar Orbiter will provide additional information on this question.
When studying total precipitation fractions (Section \ref{subsec:indiv_event}) we assumed $\lambda$=0.1 AU, similar to  typical mean free paths that have been derived by fits of ground level enhancement measurements, due to $>$500 MeV protons (e.g. $\lambda =0.27$ AU used by \cite{Bieber_2002}). 

Overall, even in the presence of scattering, back-precipitation is generally highly inefficient, with instantaneous precipitation fractions being below 2\% in our simulations. $P$ decreases strongly with height of injection, $r_i$ (Figure \ref{percent_vs_inject_rad_plot}) because of this proton injection at large radial distances cannot meaningfully extend the precipitation on the solar surface. %, e.g. with a radial dependence of $\sim r_i^-???$ for a scattering mean free path $\lambda = 0.1$ AU.

 It is also possible to use the results of our simulations, specifically the radial dependence of instantaneous precipitation, to estimate an upper limit $\overline P$ to the total precipitation fraction within the CME shock scenario. Using a variety of idealised injection functions we have shown that when the acceleration takes place over a broad range of radial distances, for example in the case of very fast shocks, lower values of $\overline P$ are obtained. When $\overline{P}$ was calculated for eight solar eruptive events that resulted in LDGRFs the values obtained range from $\sim 0.56$\% to $\sim 0.93$\%, with the smallest values corresponding to events with the fastest shocks and longer durations. This is because the shocks for these events spend less time close to the solar surface, where the precipitation is efficient. The radial position of initial particle injection ($r_{ini}$) has a substantial effect on the total precipitation fraction.
All the events analysed were associated with fast CME-driven shocks: a shock with the average speed of our subset of events could reach  70 R$_\odot$ in less than 6.7 hours. The fastest shock would cover this distance in $\sim$ 5 hours.

\cite{Den2019} used observations to directly compare the number of protons interacting at the Sun, $N_\mathit{LDGRF}$, with the number of SEP protons at 1 AU, $N_\mathit{SEP}$ . From $N_\mathit{LDGRF}$ and $N_\mathit{SEP}$  they calculated a lower limit on the total precipitation fraction required for the validity of the CME shock scenario, in which the two populations have a common origin. They found that precipitation fractions greater than 10\% were required in the majority of the 14 events considered, as shown by their Figure 8.
For the 2011 March 7 and 2012 January 23 events they reported that a $>$90\% precipitation fraction is required by the CME scenario. Our modelling of the same events found values more than two orders of magnitude smaller, with $\overline P$ $\sim$ 0.60 \% and $\sim$ 0.67 \%, respectively.
The small value ($\sim 2.9$\%) they obtained  for the 2014 February 25 event is approximately a factor of 4.5 greater than our estimate ($\sim 0.64$\%). In the case of the 2012 March 7-10 events, since it was not possible to evaluate the individual contributions in terms of SEP intensities at 1 AU, they provided a single precipitation fraction value of $\sim 18$\%, which exceeds by at least 19 times the values that we calculated for three CMEs on 2012 March 7 - 9 considered in this work ($\sim 0.56$ - $0.93$\%); similar results are expected for the 2012 March 10 CME, characterised by an intermediate speed. 
Overall we conclude that while in many events the direct observational comparison between the interacting and SEP populations gives rise to a requirement of large precipitation fractions under the assumption of CME shock acceleration of both populations \citep{Den2019}, for the same events our modelling cannot produce them due to the strong effect of magnetic mirroring. This poses a problem for the CME shock hypothesis for LDGRFs.

% How could larger precipitation fractions be obtained from our model for the eight LDGRF events?

 In order to obtain larger precipitation fractions from our model for the eight LDGRF events we would have to choose the injection function with fastest rise and shortest decay phase (i.e. tending towards a delta function injection at the radial position of peak injection - an instantaneous injection), but the result of this would be a reduction in the overall duration of precipitation. Even with this choice of injection function $\overline P$ would overall still remain smaller than 1.5\% (for our peak injection at 5.0 R$_\odot$, see Table \ref{sim-table}).

% In summary, for the CME-driven shock to be the source of the energetic protons causing the $\gamma$ ray emission in LDGRFs, observations show that back-precipitation needs to be extended in time \citep{Sha2018} and characterised by large total precipitation fractions \citep{Den2019}. However, our simulations show that while short lived injections do produce higher precipitation fractions, they cannot account for the long duration of the emission. 
% For specific LDGRF events our model calculations produce precipitation fractions much smaller than those required to reconcile $\gamma$ ray and SEP proton observations via the CME-driven shock scenario \citep{Den2019}. For two LDGRF events (2011-03-07 and 2012-01-23) they are two orders of magnitude smaller.

 Our simulations also show that with increasing radial distance of injection the emission region on the solar surface moves westwards. Particles that precipitate early in time after injection tend to follow Parker spiral magnetic field lines, but with increasing propagation time the protons deviate due to drifts and finite Larmor radius effects. This deviation becomes larger with increasing particle energy (Figure \ref{lat-vs-long-20Rsun-0_1au-plot}).
 
Our estimates of the proton precipitation count rate due to the first shock in the 2012 March 7 event (Figure \ref{10min-hists}) decay much faster than even the comparatively rapid early decay of the observed $\gamma$ ray profile, and contributions towards the precipitation quickly fall orders of magnitude indicating that $\gamma$ ray production due to the back-precipitation of energetic protons from a CME driven-shock is inconsistent with the long duration of the observed $\gamma$ ray profiles for $\lambda = 0.1$ AU.

In summary the above results present a challenge to the CME shock acceleration scenario for LDGRFs as follows:
\begin{itemize}
    \item Long after the eruptive event, CME shocks are very far from the solar surface and back-precipitation is extremely difficult. A faster CME shock only exacerbates this problem.
    
    \item  The total precipitation fraction, $\overline P$, values obtained in our study were typically smaller than 1.5\%, while work by \cite{Den2019} has indicated that in several LDGRF events a much larger value of $\overline P$ is required for the validity of the CME shock scenario.
    
    \item Time-extended acceleration and large total precipitation fractions cannot be reconciled according to our simulations. A model of an event that makes the duration of the acceleration longer will result in smaller total precipitation fractions.
    
    \item The specific shape of the precipitation count rate versus time obtained from our simulations displays a much faster  decay than that observed in  LDGRF intensity profiles (Figure \ref{10min-hists}).
    
\end{itemize}

In our simulations 300 MeV protons were considered. However, we do not expect a significant difference in total precipitation fractions for higher-energy particles as scattering depends on energy only weakly and the magnetic mirror effect does not depend on particle energy.

The shock speeds used to calculate the $\overline P$ values were based on the plane-of-the-sky CME speeds averaged over the LASCO field of view (2-32 R$_\odot$). Therefore, they underestimate the CME velocities close to the Sun and, in general, the corresponding space (3D) speeds due to projection effects, especially for events originating far from the solar limb. Consequently, derived total precipitation fractions are expected to be overestimates. However, their values remain very small even with these assumptions.

In this paper we did not consider the expansion of the magnetic field in the corona and near the photosphere and their effects on particle back-precipitation.  
The coronal magnetic field is complex and  modelling it accurately involves using either an MHD simulation or a potential field source surface  model. These types of simulations show significant expansion of the magnetic field, which provides an increased challenge to charged particles that attempt to propagate deep into the solar atmosphere.
Below the corona the magnetic field penetrates the photosphere at the edges of the convective cells and forms a `canopy' at the base of the corona, where magnetic pressure dominates \citep{Seck_1991,Seck_1992}. The result of the magnetic field being swept to the edges of convective cells is an inhomogeneous magnetic field, with flux `bundles' having magnetic field strengths of the order of $10^3$ Gauss, while the average magnetic field strength of the photosphere is of the order of a few Gauss \citep{Seck_1991}. 
Therefore, particles propagating through coronal and photospheric magnetic fields will experience increased mirroring, not considered in this study. Including these effects will reduce the precipitation fractions further due to the large mirror ratio associated with the inhomogeneous magnetic field structure. We plan to address this question in future work.

Further work is required to fully understand the contribution of the CME shock scenario to the production of $\gamma$ rays during LDGRFs. Modelling events in detail, especially the complex events that are more difficult to explain using the shock source scenario, such as the 2012 March 7 event, would provide a clearer picture of its contribution. In further work we will examine the effects of including the Heliospheric current sheet and current sheets in the vicinity of the CME in future modelling.
%In order to assess whether drift motion along the current sheet plays a significant role in the back-precipitation process, in future work we will examine the effects of including the Heliospheric Current Sheet (HCS) in further modelling. It is possible that the HCS, where the adiabatic motion of charged particles breaks down, may play a significant role in the propagation of the protons from large distances back to the solar surface. There have been observations of a CME-trailing current sheet in the 2017 September 10 event, as reported by \cite{Koc_2020}, which may also contribute to a more localised precipitation with a higher total precipitation fraction.

\vspace{1em}% just leaves a line
\noindent {\bf Acknowledgments}

A. Hutchinson, S. Dalla and T. Laitinen acknowledge support from the UK Science and Technology Facilities Council (STFC), through a Doctoral Training grant - ST/T506011/1 and grants ST/R000425/1 and ST/V000934/1. 
C.O.G. Waterfall and S. Dalla acknowledge support from NERC grant NE/V002864/1.
G.A. de Nolfo and A. Bruno are supported by the NASA Goddard Space Flight Center / Internal Scientist Funding Model (ISFM) grant HISFM18.

This work was performed using resources provided by the Cambridge Service for Data Driven Discovery (CSD3) operated by the University of Cambridge Research Computing Service (www.csd3.cam.ac.uk), provided by Dell EMC and Intel using Tier-2 funding from the Engineering and Physical Sciences Research Council (capital grant EP/P020259/1), and DiRAC funding from the Science and Technology Facilities Council (www.dirac.ac.uk).

\vspace{5mm}

\bibliographystyle{aa}
\bibliography{ldgrf_biblio}

%% the collaboration and author truncation com
%mands are used.  It has to


\end{document}